\documentclass[acmsmall, nonacm, authorversion]{acmart}

\usepackage{thm-restate}
\usepackage[algoruled,vlined,linesnumbered,titlenotnumbered,noend]{algorithm2e}

\SetKwBlock{Let}{let}{}
\SetCommentSty{mycommfont}
\DontPrintSemicolon 

\usepackage{pifont,bbding,stmaryrd,bm} 

\usepackage{xcolor}
\usepackage{pgfmath}
\usepackage{wrapfig} 
\usepackage{xspace} 
\usepackage[capitalise]{cleveref}
\usepackage{tikz}
\usetikzlibrary{calc,positioning,backgrounds,decorations.pathmorphing,shapes,calc,arrows.meta,graphs}
\usepackage{tikz-layers}
\usetikzlibrary {arrows.meta}

\usepackage{pgfplots}
\usepackage{tabularray}

\usepackage{lineno}

\usepackage{caption}

\let\oldnl\nl
\newcommand{\nonl}{\renewcommand{\nl}{\let\nl\oldnl}}

\tikzstyle{irnode}=[font=\Large]
\tikzstyle{bnfnode}=[font=\Large]
\colorlet{eventcolor}{green!50!black}
\colorlet{msgcolor}{red!70!black}
\colorlet{typecolor}{blue!80!black}
\colorlet{keycolor}{green!50!black}
\colorlet{valcolor}{red}
\colorlet{attrcolor}{blue}
\colorlet{opcolor}{eventcolor}
\colorlet{procedurenamecolor}{green!50!black}
\colorlet{relcolor}{blue}
\definecolor{varcolor}{RGB}{166,42,42} 
\definecolor{proccolor}{rgb}{0.82, 0.1, 0.26} 
\definecolor{threadcolor}{rgb}{0.82, 0.1, 0.26} 

\newcommand\hide[1]{}
\newcommand\true{\mathsf{true}}
\newcommand\false{\mathsf{false}}
\newcommand{\tuple}[1]{\left\langle#1\right\rangle}

\newcommand\of[2]{{#1}\left(#2\right)}
\newcommand\dof[3]{{#1}\left(#2\right)\left(#3\right)}

\newcommand\ii{i}
\newcommand\jj{j}

\newcommand\transcup\Cup

\newcommand\height{{\mathsf{height}}}

\newcommand\reldelete\ominus
\newcommand\compose\circ

\newcommand\undf\bot

\newcommand\set[1]{\left\{{#1}\right\}}
\newcommand\setcomp[2]{\left\{{#1}\mid\,{#2}\right\}}
\newcommand\addtoset\oplus
\newcommand\nat{{\mathbb N}}

\newcommand\assigned{:=}

\newcommand\ordering\preceq

\newcommand\sizeof[1]{\left|#1\right|}

\newcommand\emptyword\epsilon

\newcommand\wposition[2]{{#1}[{#2}]}

\newcommand\wtranspose\otriangle
\newcommand\lengthof[1]{\left|#1\right|}
\newcommand\app\bullet

\newcommand\wfilter\odot

\newcommand\valset{\mathcal{V}}
\newcommand\val{\mathsf{v}}

\newcommand\tmpvar{\mathsf{tmp}}

\newcommand\prog{{\mathcal P}}

\newcommand\thread\theta
\newcommand\pthread\phi
\newcommand\sthread\psi
\newcommand\ethread\eta

\newcommand\threadset\Theta
\newcommand\lbl\lambda

\newcommand\lblset\Lambda

\newcommand\attrtype[1]{\textcolor{attrcolor}{\mathsf{#1}}}
\newcommand\eventattr{{\attrtype{event}}}

\makeatletter
\newcommand\mmovesto[2][]{\ext@arrow 0359\Rightarrowfill@{#1}{#2}}
\makeatother

\newcommand\run\rho

\newcommand\pth\pi

\newcommand\conf\gamma

\newcommand\transitionset\Delta
\newcommand\op{\textcolor{opcolor}{\mathsf{op}}}
\newcommand\iop[1]{\textcolor{opcolor}{\op_{#1}}}

\newcommand\opseq\rho
\newcommand\emptyopseq\epsilon

\newcommand\codespace{; \hspace{5pt}}

\SetKwProg{Let}{let}{ in}{} 

\SetKwFunction{Return}{\bf return} 
\SetKwFunction{Exit}{\bf exit} 
    \SetKwInOut{Input}{Input}
    \SetKwInOut{Output}{Output}

\newcommand\event{{\color{eventcolor}{e}}}

\newcommand\ievent[1]{{\color{eventcolor}{\event_{#1}}}}

\newcommand\valattr{{\color{attrcolor}{\mathsf{val}}}}
\newcommand\valattrof[1]{{#1}\mathop{\cdot}\valattr}

\newcommand\typeattr{{\color{attrcolor}{\mathsf{type}}}}
\newcommand\typeattrof[1]{{#1}\mathop{\cdot}\typeattr}

\newcommand\history{h}

\newcommand\ihistory[1]{\history_{#1}}

\newcommand\rfrelsof[1]{{\color{relcolor}{\mathsf{{RF}}}}}

\newcommand\corelsof[1]{{\color{relcolor}{\mathsf{{CO}}}}}

\newcommand\frrelsof[1]{{\color{relcolor}{\mathsf{{FR}}}}}

\newcommand\concurrentwith\parallel

\newcommand\interval\alpha
\newcommand\binterval\beta
\newcommand\intervals\gamma
\newcommand\mkinterval[2]{
  \left[{#1},{#2}\right]}

\newcommand\rightattr{{\mathsf{right}}}
\newcommand\rightattrof[1]{{#1}\mathop{\cdot}\rightattr}

\newcommand\leftattr{{\mathsf{left}}}
\newcommand\leftattrof[1]{{#1}\mathop{\cdot}\leftattr}

\newcommand\pushop{\textcolor{opcolor}{\mathsf{push}\!}}
\newcommand\pushopof[1]{\textcolor{opcolor}{!#1}}

\newcommand\popop{\textcolor{opcolor}{\mathsf{pop}\!}}
\newcommand\popopof[1]{\textcolor{opcolor}{?{#1}}}

\newcommand\sbot{\bot}
\newcommand\eventattrof[1]{{#1}\mathop{\cdot}\eventattr}
\newcommand\callattr{{\mathsf{call}}}
\newcommand\callattrof[1]{{#1}\mathop{\cdot}\callattr}
\newcommand\returnattr{{\mathsf{ret}}}
\newcommand\returnattrof[1]{{#1}\mathop{\cdot}\returnattr}
\newcommand\retattr{{\mathsf{ret}}}
\newcommand\retattrof[1]{{#1}\mathop{\cdot}\retattr}

\newcommand\intrvattr{{\mathsf{intrv}}}
\newcommand\intrvattrof[2]{{#1}\mathop{\cdot}\intrvattr}

\newcommand\evsetattr{{\mathsf{evs}}}
\newcommand\evsetattrof[1]{{#1}\mathop{\cdot}\evsetattr}

\newcommand\pushattr{{\mathsf{push}}}

\newcommand\popattr{{\mathsf{pop}}}

\newcommand\pushcallattrof[1]{{#1}\mathop{\cdot}\pushattr\mathop{\cdot}\callattr}
\newcommand\pushretattrof[1]{{#1}\mathop{\cdot}\pushattr\mathop{\cdot}\retattr}
\newcommand\popcallattrof[1]{{#1}\mathop{\cdot}\popattr\mathop{\cdot}\callattr}
\newcommand\popretattrof[1]{{#1}\mathop{\cdot}\popattr\mathop{\cdot}\retattr}
\newcommand\intervalattr{{\mathsf{interval}}}
\newcommand\intervalattrof[1]{{#1}\mathop{\cdot}\intervalattr}

\newcommand\tattr{{\mathsf{T}}}
\newcommand\tattrof[1]{{#1}\mathop{\cdot}\tattr}
\newcommand\iattr{{\mathsf{I}}}
\newcommand\iattrof[1]{{#1}\mathop{\cdot}\iattr}

\newcommand\lsubattr{{\mathsf{lsub}}}
\newcommand\lsubattrof[1]{{#1}\mathop{\cdot}\lsubattr}
\newcommand\rsubattr{{\mathsf{rsub}}}
\newcommand\rsubattrof[1]{{#1}\mathop{\cdot}\rsubattr}
\newcommand\lkeyattr{{\mathsf{lkey}}}
\newcommand\lkeyattrof[1]{{#1}\mathop{\cdot}\lkeyattr}
\newcommand\hkeyattr{{\mathsf{hkey}}}
\newcommand\hkeyattrof[1]{{#1}\mathop{\cdot}\hkeyattr}
\newcommand\rkeyattr{{\mathsf{rkey}}}
\newcommand\rkeyattrof[1]{{#1}\mathop{\cdot}\rkeyattr}

\newcommand\hprecedess[1]{\prec_{#1}}
\newcommand\hconcurrent[1]{\,\Box_{#1}\,}
\newcommand\hrefines\sqsubseteq

\newcommand\trace\tau
\newcommand\ptrace{\trace'}
\newcommand\itrace[1]{\trace_{#1}}
\newcommand\adt{{\mathsf{ADT}}}
\newcommand\stackadt{{\mathsf{Stack}}}
\newcommand\queueadt{{\mathsf{Queue}}}
\newcommand\denotationof[1]{\llbracket{#1}\rrbracket}
\newcommand\collapse{{\mathsf{collapse}}\!}
\newcommand\collapseof[1]{\of\collapse{#1}}
\newcommand\linset{{\mathsf{Lin}}\!}
\newcommand\linsetof[1]{\of\linset{#1}}

\newcommand\popempty{{\mathsf{PopEmpty}}}
\newcommand\popemptyof[1]{\of\popempty{#1}}

\newcommand\intervalintersects\boxtimes

\SetKwFunction{Sort}{\bf sort} 

\newcommand\extremevar{{\mathsf{E}}}
\newcommand\crowdedvar{{\mathsf{Populated}}}
\newcommand\desertedvar{{\mathsf{Deserted}}}
\newcommand\valvar{{\mathsf{V}}}

\newcommand\currentvar{{\mathsf{current}}}
\newcommand\listvar{\ell}
\newcommand\flagvar{{\mathsf{flag}}}
\newcommand\internalvar{{\mathsf{Internal}}}

\newcommand\csegmentsalg{{\mathsf{PSegments}}}
\newcommand\csegmentsalgof[1]{\of\csegmentsalg{#1}}
\newcommand\dsegmentsalg{{\mathsf{DSegments}}}
\newcommand\dsegmentsalgof[1]{\of\dsegmentsalg{#1}}
\newcommand\linearizablealg{{\mathsf{Linearizable}}}
\newcommand\linearizablealgof[1]{\of\linearizablealg{#1}}
\newcommand\partitionalg{{\mathsf{Partition}}}
\newcommand\partitionalgof[2]{\of\partitionalg{#1,#2}}

\newcommand\sortalg{{\mathsf{Sort}}}
\newcommand\sortalgof[1]{\of\sortalg{#1}}
\newcommand\extremealg{{\mathsf{Extreme}}}
\newcommand\extremealgof[2]{\of\extremealg{#1,#2}}

\newcommand\searchalg{{\mathsf{Search}}}
\newcommand\searchalgof[2]{\dof\searchalg{#1}{#2}}

\newcommand\buildpqtree{{\mathsf{PQTree}}}
\newcommand\buildpqtreeof[1]{\of\buildpqtree{#1}}

\newcommand\completion{{\mathsf{Complete}}}
\newcommand\completionof[1]{\of\completion{#1}}

\newcommand\optoval{{\mathsf{OpToVal}}}
\newcommand\optovalof[1]{\of\optoval{#1}}

\newcommand\violin{\sc{Violin}\xspace}
\newcommand\limo{\sc{LiMo}\xspace}

\newcommand\gnode{p} 
\newcommand\ignode[1]{\gnode_{#1}} 
\newcommand\nullnode{{\mathsf{null}}}
\newcommand\bigo{{\mathcal{O}}}

\newcommand\hempty[1]{\mathtt{hasEmpty(#1)}}
\newcommand\match{\mathtt{matched}}
\newcommand\data{\mathcal{D}}

\newcommand{\proj}[2]{\texttt{proj(}\ensuremath{#1}\texttt{,}\ensuremath{#2}\texttt{)}}

\newcommand\intv[1]{\left[#1\right]}

\newcommand\tree{\mathtt{T}}
\newcommand\T{\mathbb{T}}

\newcommand\ctn[1]{\mathtt{contains(#1)}}

\begin{document}

\title{Efficient Linearizability Monitoring}

\author{Parosh Aziz Abdulla}
\orcid{0000-0001-6832-6611}
\affiliation{%
  \institution{Uppsala University}
  \city{Uppsala}
  \country{Sweden}
}
\affiliation{%
  \institution{Mälardalen University}
  \city{Västerås}
  \country{Sweden}
}
\email{parosh@it.uu.se}

\author{Samuel Grahn}
\orcid{0009-0004-1762-8061}
\affiliation{%
  \institution{Uppsala University}
  \city{Uppsala}
  \country{Sweden}
}
\email{samuel.grahn@it.uu.se}

\author{Bengt Jonsson}
\orcid{0000-0001-7897-601X}
\affiliation{%
  \institution{Uppsala University}
  \city{Uppsala}
  \country{Sweden}
}
\email{bengt@it.uu.se}

\author{S. Krishna}
\orcid{0000-0003-0925-398X}
\affiliation{%
  \institution{IIT Bombay}
  \city{Mumbai}
  \country{India}
}
\email{krishnas@cse.iitb.ac.in}

\author{Om Swostik Mishra}
\orcid{0009-0001-6858-6605}
\affiliation{%
  \institution{IIT Bombay}
  \city{Mumbai}
  \country{India}
}
\email{21b090022@iitb.ac.in}

\begin{abstract}
This paper revisits the fundamental problem of monitoring the linearizability of concurrent stacks, queues, sets, and multisets.
Given a history of a library implementing one of these abstract data types, the monitoring problem is to answer whether the given history is linearizable.
For stacks, queues, and (multi)sets,  we present monitoring algorithms with complexities $\bigo(n^2)$, $\bigo(n\; log\, n)$, and $\bigo{(n)}$, respectively, where $n$ is the number of operations in the input history.  
For stacks and queues, our results hold under the standard assumption of {\it data-independence}, i.e., the behavior of the library is not sensitive to the actual values stored in the data structure.
Past works to solve the same problems have cubic time complexity and (more seriously) have correctness issues: they either (i) lack correctness proofs or (ii)  the suggested  correctness proofs are erroneous (we present counter-examples), or (iii) have incorrect algorithms.
Our improved complexity results rely on substantially different algorithms for which we provide detailed proofs of correctness.
We have implemented our stack and queue algorithms in $\limo$ (Linearizability Monitor). We evaluate $\limo$ and compare it with 
the state-of-the-art tool $\violin$ -- whose correctness proofs we have found errors in -- which checks for linearizability violations. Our experimental evaluation confirms that $\limo$ outperforms $\violin$ regarding both efficiency and scalability.
\end{abstract}

\maketitle
\section{Introduction}
\label{introduction:section}

In concurrent programs, inter-thread communication is often performed using libraries that implement shared data structures, specified as abstract data types $\adt$s, such as stacks, queues, sets, counters, snapshots, etc.
Such a library, together with its clients, can be considered as
a collection of threads running concurrently and communicating through the shared data structure.
Each thread executes a sequence of operations on the library,
issuing calls (invocations) followed by receiving the associated responses.
Traditionally, we refer to an execution of the system as a {\it history}.
The design of concurrent libraries implementing $\adt$s is tricky.
On the one hand, thread synchronization should be minimal to decrease communication and increase throughput.
On the other hand, we need sufficient synchronization to guarantee the correctness of the library.
Given the ubiquity of concurrent data structure libraries and the intricacy of their designs, it is vital to develop methods and frameworks to increase confidence in their correctness.
%
Linearizability \cite{HeWi:linearizability} is the standard correctness criterion for concurrent libraries.
It states that each library operation can be considered to be performed atomically at some point, called the linearization point, between its invocation and return.
In this manner, linearizability makes each operation appear to ``take effect'' instantaneously while it preserves the order of non-concurrent operations.
Linearizability allows client threads to understand the data structure in terms of atomic actions without considering the complications of concurrency.
Since linearizability is the accepted correctness criterion for concurrent libraries, proving the latter's correctness amounts to proving their linearizability.
Static verification techniques for proving linearizability, such as model checking and theorem proving, suffer from scalability issues due to the complicated nature of library implementations typically involving unbounded data domains such as integers and counters and comprising dynamically evolving heaps \cite{AbdullaJT18}.
Due to these practical limitations of static methods, many researchers have developed complementary, dynamic techniques like testing and monitoring.
\smallskip 

In monitoring, we analyze the linearizability of only a {\it single} library history rather than  {\it all} histories, which is the case for static verification frameworks.
The goal is to obtain scalability by sacrificing completeness.
In monitoring, we run the program, record a {\it single} history that it generates, and then check whether the recorded history is linearizable.
In particular, we do not need to consider the underlying library implementation and avoid analyzing its complex state space.
Monitoring offers several advantages.
First, like other dynamic techniques, it increases confidence in the correctness of the implementation and provides eager detection of bugs that can lie deep in the program's state space \cite{DBLP:journals/pacmpl/EmmiE18}.
Second, monitoring is scalable, allowing the analysis of histories with thousands of events.
While incomplete, analyzing long histories still gives extensive state space coverage.
Since we would like to analyze long histories, a monitoring algorithm's time complexity must be low in the length of the input history.
Linear time complexity is satisfied by several monitors designed for checking typical safety properties such as assertion validation \cite{DBLP:journals/sttt/HavelundR04}.
In contrast, monitoring linearizability against arbitrary $\adt$s requires exponential time \cite{DBLP:journals/siamcomp/GibbonsK97}.
This high complexity rules out practically scalable monitoring algorithms that cover general $\adt$ classes.
%

One way to avoid the exponential time complexity is to develop algorithms for specific useful $\adt$s such as stacks, queues, and sets.
For instance, \cite{DBLP:journals/pacmpl/EmmiE18} introduced a class of $\adt$s, called {\it collection types}, that cover common $\adt$s provided by libraries, e.g., {\tt java.util.concurrent}, such as stacks, queues, sets, and maps.
Other papers that provide polynomial monitoring algorithms include \cite{DBLP:conf/pldi/EmmiEH15,DBLP:conf/vmcai/PetersonCD21}.
While the problem seems simple, and the algorithms described in the above papers are elegant and efficient, the problem is more complex.
Unfortunately, in the case of stacks and queues, previous polynomial algorithms either lack a correctness proof \cite{DBLP:conf/pldi/EmmiEH15} (and the proof seems far from trivial), or the main correctness lemmas are erroneous \cite{DBLP:journals/pacmpl/EmmiE18} (and how to fix them is not apparent to us), or the algorithms return incorrect responses for some inputs \cite{DBLP:conf/vmcai/PetersonCD21}.
Thus, to our knowledge, polynomial time algorithms for monitoring stacks, queues, sets, or multisets have yet to be proven correct.

%
\smallskip 

This paper introduces monitoring algorithms
with complexities $\bigo(n^2)$, $\bigo(n\; log\, n)$, and $\bigo(n)$, respectively, for stacks, queues, and (multi)sets, where $n$ is the length of the input history.
Our results for stacks and queues hold under the standard assumption of {\it data-independence}, i.e., the behavior of the library is not sensitive to the actual values stored in the stack or the queue \cite{Integrated:tacas13,BouajjaniEGH17,DBLP:conf/popl/Wolper86}.
On the other hand, it is not realistic to assume data independence for set and multiset algorithms, and we do not assume that either in our algorithms.
We refer to \cref{overview:data:independence:subsubsection} for more details about data-independence.
The algorithms we provide for the different data structures use various and mutually different strategies.
We provide a detailed overview of the main concepts and ideas in \cref{overview:section}.
We decompose the structure of a linearization for each of the $\adt$s considered. We note that the structure can be inductively defined for stacks, and we develop a recursive algorithm that matches this structure. For queues, we note that every unlinearizable history must have a \emph{critical pair}, i.e., a set of two values of the history, the operations on which are unlinearizable. Finally, for sets and multisets, we note that it is sufficient to count certain events and compare the counters to decide linearizability; the construction of the linearization can be done greedily if these checks are successful. We provide correctness proofs for all our algorithms in the appendix.
\smallskip 

We have also implemented a tool, $\limo$ (Linearizability Monitor), based on our algorithm. We evaluate it experimentally, demonstrating its efficiency and scalability. Apart from the correctness proofs, the experimental results also validate the complexity analysis of our algorithm. We also compare it with the tool, Violin, which checks for linearizability violations \cite{DBLP:conf/pldi/EmmiEH15}. While the suggested  formal proof \cite{DBLP:journals/pacmpl/EmmiE18} for the tool $\violin$ does not hold,  we are yet to find a counterexample to the tool. $\limo$ significantly outperforms $\violin$ on a massive set of benchmarks.

\section{Related Work}
As mentioned in \cref{introduction:section}, several previous problematic approaches to solving the linearizability problem have appeared in recent years. First, we discuss these and their issues.
Then, we will mention more generally related works.

\subsection{Issues with Closely Related Works}
\label{sec:related:issues}
%
We start by describing the issue with the algorithm presented in \cite{DBLP:conf/vmcai/PetersonCD21}. In it, the authors present an algorithm for monitoring linearizability in quadratic time. Their algorithm is highly efficient, and has excellent parallelization using GPUs. However, the algorithm incorrectly identifies a linearizable history as unlinearizable.  The history can be seen in the full version.  
This issue has been confirmed by the authors, who have concluded that a fix to this issue would increase the complexity from $\bigo(n^{2})$ to $\bigo(n^{3})$.

The second work which is closely related to ours is \cite{DBLP:journals/pacmpl/EmmiE18}, which  specializes into a class of \emph{collection types} and does a  good analysis of some commonly occurring patterns in data structures. Moreover, their implementation, as presented in \cite{DBLP:conf/pldi/EmmiEH15}, is an elegant algorithm, and the first polynomial algorithm for monitoring linearizability for concurrent structures.
Unfortunately, this algorithm lacks a formal correctness proof. While \cite{DBLP:journals/pacmpl/EmmiE18}  tries to establish the correctness of the algorithm in \cite{DBLP:conf/pldi/EmmiEH15},   one of the main lemmas in \cite{DBLP:journals/pacmpl/EmmiE18} which is used to argue the correctness of the algorithm  
does not hold,   specifically, Lemma 5.9.  This lemma is one part of their proof that the linearizability of a history can be encoded as a Horn formula. According to \cite{DBLP:journals/pacmpl/EmmiE18}, 
this construction argues for the correctness of the algorithm presented in \cite{DBLP:conf/pldi/EmmiEH15}, which does not have any other formal proof.

\smallskip 

Lemma 5.9 (\cite{DBLP:journals/pacmpl/EmmiE18} ) states that, for any \emph{ground}, \emph{quantifier-free} \emph{operation-order formula} $\varphi$ in CNF, with total order axioms, the \emph{Hornification} $\bar{\varphi}$ of $\varphi$ is satisfiable if and only if $\varphi$ is.  Our counterexample to this consists of three operations, $a, b, c$, in the formula $\varphi = g \wedge totality \wedge transitivity$, where (i) $g$ is in CNF : it is a  conjunction of disjunctions of predicates of the form $x<y, \neg (x<y)$, $x, y \in \{a,b,c\}$, $x \neq y$, (ii) \emph{totality} encodes a total order between $a,b,c$ and (iii) \emph{transitivity} as the name suggests,  encodes transitivity  in the order between $a,b,c$.
While $\varphi$ can be seen in the full version,    
we state that $\varphi$ satisfies all of the requirements. First, it is ground and quantifier-free. It is an operation order formula over the ordering $<$. It is in CNF form, and it contains the total-order axioms.
Moreover, $\varphi$ is unsatisfiable.
However, the hornification $\bar{\varphi}$ is satisfied by assigning $<$ the empty relation, or $<~= \set{(a, b)}$. Since $\varphi$ only has Horn clauses $c$ of length $|c| > 1$,  there is always a negative literal in every clause in $\bar{\varphi}$, which makes the clause satisfiable. Specifically, the hornification of the totality axioms do not enforce totality. 
This counterexample to Lemma 5.9 of \cite{DBLP:journals/pacmpl/EmmiE18} does not immediately produce a counterexample to the algorithm presented in \cite{DBLP:conf/pldi/EmmiEH15}. It only establishes that the proposed proof in  \cite{DBLP:journals/pacmpl/EmmiE18} for the algorithm in \cite{DBLP:conf/pldi/EmmiEH15} is incorrect.
We have not found a counterexample to the algorithm; however,
there is no formal proof for the algorithm of \cite{DBLP:conf/pldi/EmmiEH15}.

\subsection{Other Related Work}
Extensive work has been done on developing static verification techniques for proving the linearizability of concurrent programs.
A non-exhaustive list includes
\cite{Integrated:tacas13,DBLP:conf/cav/AmitRRSY07,DBLP:conf/popl/BouajjaniEEH15,DBLP:conf/popl/DoddsHK15,DBLP:conf/cav/DragoiGH13,DBLP:conf/concur/HenzingerSV13,HeWi:linearizability,DBLP:journals/corr/KhyzhaGP16,DBLP:conf/pldi/LiangF13,DBLP:conf/podc/OHearnRVYY10,DBLP:conf/pldi/SergeyNB15,DBLP:conf/oopsla/ShachamBASVY11,DBLP:conf/cav/Vafeiadis10,DBLP:conf/icse/Zhang11a}.
Several works perform monitoring in exponential time for general $\adt$s.
Early work produced an exponential-time algorithm \cite{WING1993164}, which was later improved in \cite{DBLP:journals/corr/HornK15a,lowe2017,DBLP:conf/pldi/BurckhardtDMT10}, still with exponential complexity. 
Some recent works on general concurrent objects have focused on heuristics, strategies, and common pitfalls of implementations, to alleviate the exponential complexity and achieve better real-world performance \cite{jia2023verilin}.
From the complexity theoretical point of view, \cite{DBLP:journals/siamcomp/GibbonsK97} showed that monitoring even the single-value register type is NP-hard. 
Recently, after the notification of our paper,  there has been an arxiv submission \cite{arxiv25} with similar results to ours.

\section{Overview}
\label{overview:section}
In this section, we give an overview of our algorithms.
We will first present a set of definitions and concepts that we will use to formalize the problems we are considering and the algorithms we plan to use to solve them.
We will instantiate the concepts for stacks and then extend them to the other abstract data types.
\subsection{Stacks}
\label{overview:stack:subsubsection}
\begin{figure}
\includegraphics[scale=1]{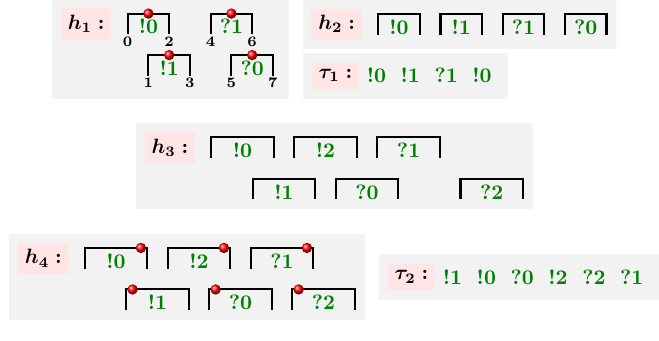}
\caption{The history $\ihistory1$ is linearizable, according to the linearization policy depicted by the red balls.
The history $\ihistory2$ is sequential and results from the linearization of $\ihistory1$.
The trace $\itrace1$ is the result of collapsing $\ihistory2$.
Notice that $\itrace1\models\stackadt$.
The history $\ihistory3$ is not linearizable.
The history $\ihistory4$ is linearizable, as witnessed by the shown linearization policy.
The trace $\itrace2$ is the result of the linearization and then collapsing $\ihistory4$.
Notice that $\itrace2\models\stackadt$.}
\label{overview:histories:fig}
\end{figure}

\subsubsection{Events and Traces}
A {\it (stack) event} $\event$ is of one of the form $\pushopof\val$, $\popopof\val$, or $\popopof\sbot$ where $\val\in\nat$ is a natural number and $\sbot\not\in\nat$ is a special value indicating that the stack is empty.
The events have their standard interpretations, i.e., pushing $i$ to the top of the stack, removing $\val\in\nat$ from the top of the stack if $\val$ is the top stack element, and returning the particular value $\sbot$ when popping from an empty stack.
We define two attributes of an event $\event$.
More precisely, we define its type attribute $\typeattrof\event:=\pushop$ if $\event$ is of the form $\pushopof\val$, and define $\typeattrof\event:=\popop$ if $\event$ is of the form $\popopof\val$ or of form $\popopof\sbot$.
We also define its value attribute $\valattrof\event:=\val$ if $\event$ is of the form $\pushopof\val$ or $\popopof\val$, and define $\valattrof\event:=\sbot$ if $\event$ is of the form $\popopof\sbot$.

A {\it trace} $\trace$ is a finite sequence of events.
We define $\denotationof\stackadt$ to be the smallest set of traces satisfying the following conditions: (i) $\emptyword\in\denotationof\stackadt$, (ii) if $\trace\in\denotationof\stackadt$ then $\pushopof\val\app\trace\app\popopof\val\in\denotationof\stackadt$, and (iii) if $\itrace1,\itrace2\in\denotationof\stackadt$ then $\itrace1\app\itrace2\in\denotationof\stackadt$.
For a trace $\trace$, we write $\trace\models\stackadt$ to denote that $\trace\in\denotationof\stackadt$.
In \cref{overview:histories:fig}, we have both $\itrace1\models\stackadt$ and $\itrace2\models\stackadt$.

\subsubsection{Operations and Histories}
A {\it (concurrent) history} describes the non-atomic execution of a set of events.
To capture this non-atomicity, we extend the definition of an event to that of an operation.
The latter comprises  an event together with two time-stamps that are its {\it call} resp.\ {\it return} points.
Formally, an operation $\op$ is a triple $\tuple{\event,\ii,\jj}$ where $\event$ is the {\it event}, and $\ii$ and $\jj$ are its call resp.\ time-stamps.
We define the following attributes for operations:
(i)  $\eventattrof\op:=\event$ is the event in $\op$, (ii) $\callattrof\op:=\ii$ is the call time-stamp, (iii) $\returnattrof\op:=\jj$ is the return time-stamp, (iv) $\intrvattrof\op:=\mkinterval\ii\jj$ is the interval defined by the call and return time-stamps, (v)
$\typeattrof\op:=\typeattrof\event$ is the type of the event in $\op$, and (vi) $\valattrof\op:=\valattrof\event$ is its value.
A {\it history} $\history$ is a set of operations in which no time-stamp occurs more than once.
We use $\evsetattrof\history$ to denote the set of events occurring in $\history$.
In \cref{overview:histories:fig}, the history $\ihistory1$ contain four operations, namely $\ihistory1=\set{\tuple{\pushopof0,0,2},\tuple{\pushopof1,1,3},\tuple{\popopof1,4,6},\tuple{\popopof0,5,7}}$.
Given two events $\ievent1$ and $\ievent2$ occuring in a history $\history$, we say that $\ievent1$ {\it precedes} $\ievent2$ in $\history$, denoted $\ievent1\hprecedess\history\ievent2$, if the interval of the operation $\iop1$ corresponding to $\ievent1$ in $\history$ precedes the interval of the operation $\iop2$ corresponding to $\ievent2$ in $\history$, i.e., $\retattrof{\iop1}<\callattrof{\iop2}$.
We say that $\ievent1$ and $\ievent2$ are {\it concurrent} in $\history$, denoted $\ievent1\hconcurrent\history\ievent2$ if neither $\ievent1\hprecedess\history\ievent2$ nor $\ievent2\hprecedess\history\ievent1$.
In \cref{overview:histories:fig}, we have $\pushopof0\hprecedess{\ihistory1}\popopof0$, while $\pushopof0\hconcurrent{\ihistory1}{\pushopof1}$.
For histories $\ihistory1$ and $\ihistory2$, we say that $\ihistory1$ {\it refines} $\ihistory2$, denoted $\ihistory1\hrefines\ihistory2$, if (i) $\evsetattrof{\ihistory1}=\evsetattrof{\ihistory2}$, and (ii) $\ievent1\hprecedess{\ihistory1}\ievent2$ if $\ievent1\hprecedess{\ihistory2}\ievent2$.
In \cref{overview:histories:fig}, we have $\ihistory2\hrefines\ihistory1$.

We say that $\history$ is {\it sequential} if it contains no  concurrent operations.
In \cref{overview:histories:fig}, the history $\ihistory2$ is sequential while $\ihistory1$ is not.
There is a close relationship between sequential histories and traces.
We can {\it collapse} a sequential history $\history$ to a trace $\trace:=\collapseof\history$ by merging the call and return events of each operation in $\history$.
We write $\history\models\stackadt$ if $\trace\models\stackadt$ where $\trace:=\collapseof\history$.
In \cref{overview:histories:fig}, $\ihistory2\models\stackadt$ since $\itrace1\models\stackadt$ and $\itrace1:=\collapseof{\ihistory2}$. 
A history is \emph{completed} if every call has a matching return. An incomplete history has multiple possible \emph{futures}, obtained by freely choosing the return value \cite{HeWi:linearizability}. In this paper, we consider only completed histories, since when collecting histories for testing, we can simply let the pending operations finish. Moreover, if a library produces an incomplete history that has a linearizable future, that does not mean that this future is what would have happened, had we let the pending operations run to its conclusion. In fact, they may return values leading to unlinearizability. Similarly, if no future is linearizable, then any return the library produces will inevitably result in an unlinearizable history. In short, we consider only completed histories because $i)$ we can reliably produce them, and $ii)$ we do not detect fewer bugs because of it.
\subsubsection{Linearizability}
For histories $\ihistory1$  and $\ihistory2$, we say that $\ihistory2$ is a {\it linearization}  of $\ihistory1$ if (i) $\ihistory2$ is sequential, and (ii) $\ihistory2\hrefines\ihistory1$.
We use $\linsetof\history$ to denote the set of linearizations of the history $\history$.
For a history $\history$, we say that $\history$ is
 {\it linearizable} (wrt.\ $\stackadt$) if $\linsetof\history\cap\denotationof\stackadt\neq\emptyset$.
In \cref{overview:histories:fig}, $\ihistory2\in\linsetof{\ihistory1}$, i.e, $\ihistory2$ is a linearization of $\ihistory1$.
Furthermore, $\ihistory1$ is linearizable since $\ihistory2\models\stackadt$.


An instance of the {\it linearizability problem} consists of a history $\history$ and the question we want to answer is whether $\history$ is linearizable or not.

\subsubsection{Differentiation and Data Independence}
\label{overview:data:independence:subsubsection}
\begin{wrapfigure}{r}{0.5\textwidth}
\includegraphics[scale=0.6]{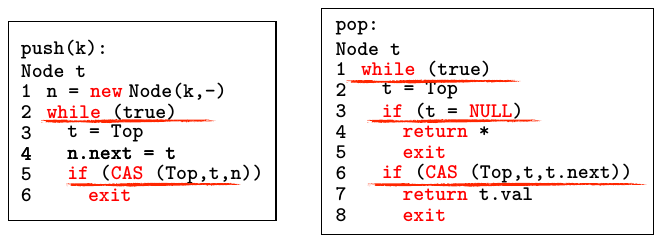}
\caption{The classical Treiber algorithm, consisting of the implementation of the push and pop operations.
The control statements in the code are underlined.
In the push operation, the control statements do not mention the input key, so it is data-independent.
The pop operation has no input data, so it is trivially data-independent.
Notice that, it is straightforward to verify data-independence through data-flow analysis; here by syntactically checking the conditions of the control statements.
}
\label{overview:Treiber:fig}
\end{wrapfigure}
It is well-known that the stack abstract data type and libraries implementing it are {\it data-independent}  (see e.g., \cite{Integrated:tacas13,DBLP:conf/popl/BouajjaniEEH15}).
Intuitively, the specification of $\stackadt$ is data-independent in the sense that if a trace belongs to $\stackadt$ then any trace that we get from $\trace$ by ``renaming values'' in $\trace$ will also be in $\stackadt$.
Consider the trace $\trace=\;\pushopof{1}\;\pushopof{0}\;\popopof{0}\;\pushopof{2}\;\pushopof{2}\;\popopof{2}\;\popopof{2}\;\popopof{1}$.
We know that $\trace\models\stackadt$.
Now, consider the trace 
$\ptrace=\;\pushopof{1}\;\pushopof{3}\;\popopof{3}\;\pushopof{5}\;\pushopof{5}\;\popopof{5}\;\popopof{5}\;\popopof{1}$.
We can derive $\ptrace$ from $\trace$ by renaming the value $2$ to $5$ and $0$ to $3$.
Notice that  $\ptrace\models\stackadt$. 
Furthermore, again as observed in \cite{Integrated:tacas13,DBLP:conf/popl/BouajjaniEEH15}, libraries implementing stacks are data-independent: the program flow is not affected by the actual values pushed or popped.
Typically such a library uses a data store, e.g., a linked list, that is organized according to the order in which the values are pushed, and not depending on the actual values that are pushed.
The library only needs to ensure that the pop operations are performed in the reverse order.
Indeed, we can verify that such libraries are data-independent by simple syntactic checks: ensure that no control statements depend on the input data (look at the classical Treiber stack algorithm in 
\cref{overview:Treiber:fig}).
It is also well-known \cite{Integrated:tacas13} that, for data-independent programs, it is sufficient to prove the correctness of the program wrt. so called {\it differentiated} histories.
We say that a trace $\trace$ (resp.\ a history $\history$) is {\it differentiated} if each operation occurs at most once in $\trace$ (resp.\ $\history$).
All the traces and histories in \cref{overview:histories:fig} are differentiated.
The trace $\pushopof1\app\pushopof2\app\popopof2\app\popopof1\app\pushopof1\app\popopof1$ is not differentiated since the event $\pushopof1$ occurs twice in the trace.
For a data-independent program $\prog$, if all the differentiated histories of $\prog$ satisfy $\stackadt$ then {\it all} histories of $\prog$ (including  the non-differentiated ones) satisfy  $\stackadt$.
Therefore, it is enough to concentrate only the differentiated traces, which is what we will do in this paper. 
For monitoring, we can handle non-differentiated executions without introducing any imprecision in the analysis.

To that end, we can wrap the library in a simple wrapper. The wrapper transforms a non-differentiated history into a differentiated one, as follows. The wrapper consists of a thread id and a thread-local counter, where the implementation of $\pushopof{x}$ culminates in a call to $\pushopof{(x, thread, counter\mathtt++)}$ of the original code. Thus, while the user sees a non-differentiated history, the monitor attached to the original code produces a differentiated history.

Note that the wrapper does not induce false positives: if the original non-differentiated history is not linearizable, then the differentiated one produced through the wrapper will also be unlinearizable. This property follows from the fact the stack and queue abstract data types are data-independent (see for example, \cite{Integrated:tacas13}). However, it can be the case that the original non-differentiated one is linearizable while the differentiated counterpart is not. However, this means that we have found an error in the library. This is good since we are finding more bugs.

\subsubsection{Value-Centric View}
\label{value:centric:subsubsection}

\begin{wrapfigure}{r}{0.6\textwidth}
\includegraphics[scale=0.6]{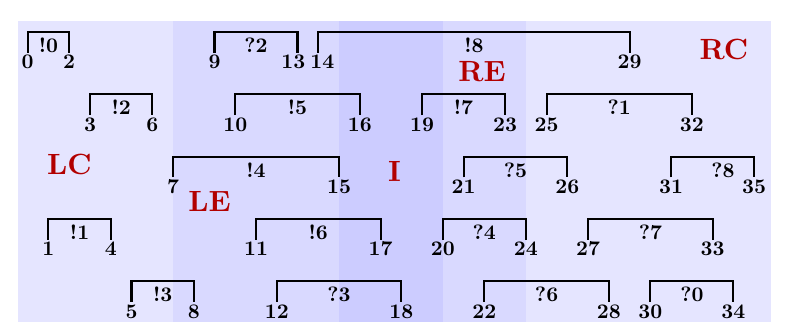}
\caption{The interval induced by the value $4$.}
\label{overview:intervals:fig}
\end{wrapfigure}

The data-independence property, and consequently the differentiation property, allows a value-centric view of histories, rather than the standard operation-centric view described above.
The value-centric view is vital for our algorithm.
Since a value $\val$ occurs at most twice in a differentiated history $\history$, namely at most in a push and a pop operation on $\val$, we can uniquely refer it as $\val$ in terms of these two operations.
We will work with {\it attributed values}, i.e., values augmented by its four time-stamps.
For instance, the {\it push-call} time-stamp of a value $\val$ is the call time-stamp of the (unique) push operation on $\val$.
We define the other three time-stamps, namely {\it push-return}, {\it pop-call}, {\it pop-return} analogously.
In the history of \cref{overview:intervals:fig}, the four time-stamps of the value $4$ are $7$, $15$, $20$, and $24$ respectively.
For any value $v$  the above time-stamps allow to partition the history into five segments.
We show the five segments for the value $4$ in \cref{overview:intervals:fig}.
The {\it Left-Context (LC)} segment is the interval  $\mkinterval07$, from the starting point of the history to the push-call time-stamp of $4$.
The {\it Left-External (LE)} segment is the interval $\mkinterval7{15}$ of the push operation of $4$, i.e., from its push-call to its push-return time-stamps.
The {\it Internal (I)} segment is the interval $\mkinterval{15}{20}$ between the push-return and pop-call time-stamps for $4$.
Intuitively, the I-segment indicates that the value $4$ is necessarily inside the stack, since its push has returned but its pop has not been called.

The {\it Right-External (RE)} segment is the dual of the LE-segment, i.e., it is the interval $\mkinterval{20}{24}$ of the pop operation of $4$, i.e., from its pop-call to its pop-return time-stamps.
The {\it Right-Context (RC)} segment is  the dual of the LC-segment,i.e., it is the interval  $\mkinterval{24}{35}$, from the pop-return time-stamp of $4$ till the ending point of the history.  
Finally, we define the {\it Total (T)} segment to be the interval $\mkinterval7{24}$, that is, from the push-call to the pop-return time-stamp of $4$.
Notice that the $T$-segment is the union of the LE-, I-, and RE-segments. In our algorithms, we use the $T$ and $I$ segments, and include the rest in this section for clarity.

Sometimes, when there is no risk for confusion, we identify operations with their types. So we write the operation $\tuple{\pushopof4,3,7}$ simply as $\pushopof4$.
In a differentiated history, this simplification will not introduce any ambiguity since the interval of an operation can be uniquely determined by its type.
For a value $\val\in\nat$, we say that $\val$ is {\it minimal} in $\history$ if there is no event $\event$ such that $\event\hprecedess\history\pushopof\val$.
We say that $\val$ is {\it maximal} in $\history$ if there is no event $\event$ such that $\popopof\val\hprecedess\history \event$.
We say that $\val$ is {\it extreme} in $\history$ if it is both minimal and maximal in $\history$.
%
Consider \cref{overview:histories:fig} again.
In $\ihistory1$ both $0$ and $1$ are extreme.
In $\ihistory3$, both $0$ and $1$ are minimal, and $2$ is maximal, while there are no extreme values.
In $\ihistory4$, the value $1$ is extreme, $0$ is minimal, and $2$ is maximal.

\subsubsection{Algorithm}
\begin{figure}
\includegraphics[scale=0.58]{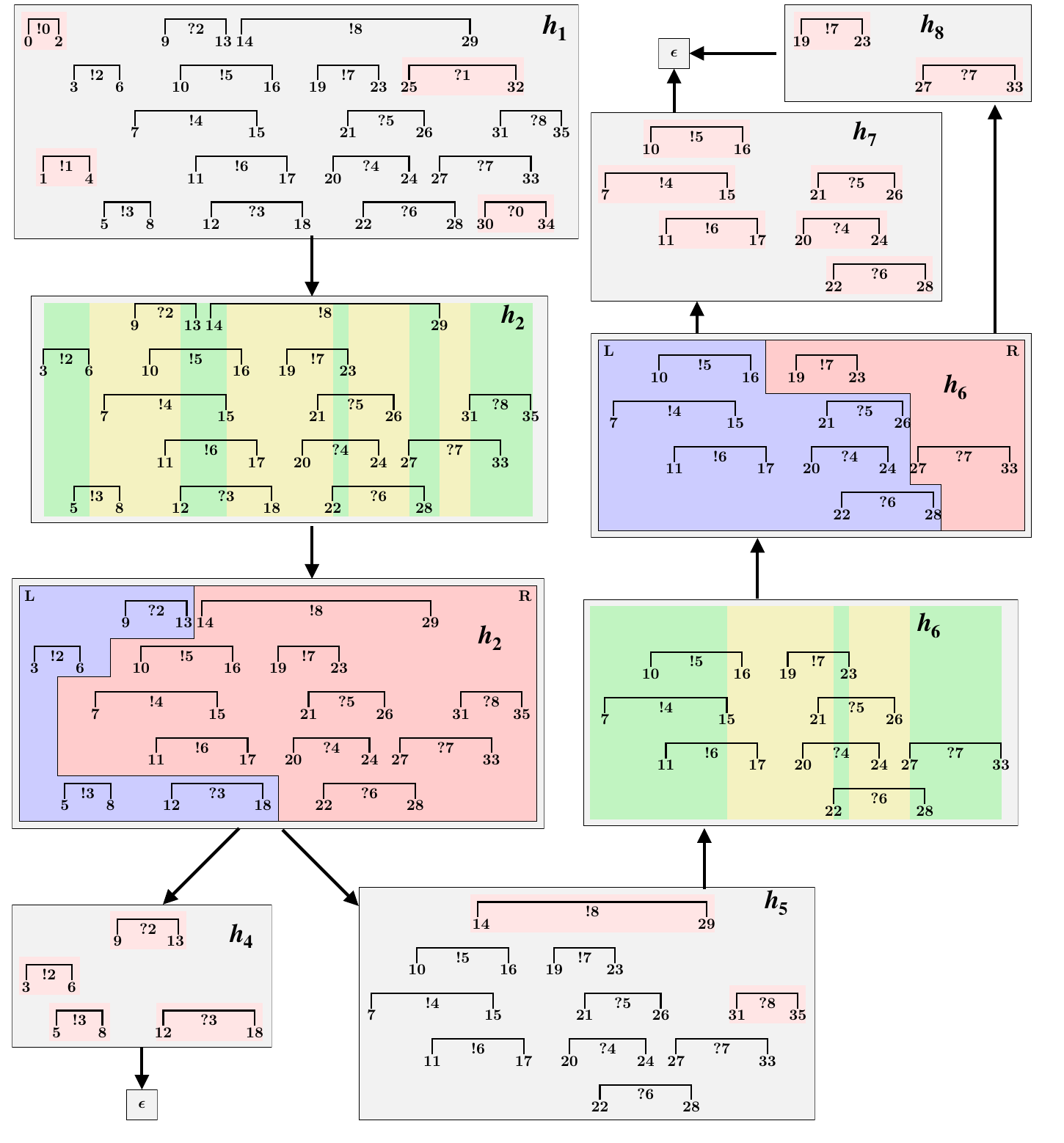}
\caption{Illustration for Stack}
\label{overview:algorithm1:fig}
\end{figure}
In \cref{overview:algorithm1:fig}, we go through the steps of our algorithm before it concludes that $\ihistory1$ (\cref{overview:algorithm1:fig}, top left) is linearizable.
The algorithm first searches for extreme values in the input history ($\ihistory1$ in this case).
If it finds any, then it will simply discard it and continue.
The reason is that the linearizability of a history $\ihistory1$, with extreme values, is equivalent to the linearizability of the (shorter) history $\ihistory2$ which we obtain by
discarding $\ihistory1$'s extreme values.
In \cref{overview:algorithm1:fig}, the values $0$ and $1$ are extreme in $\ihistory1$.
Hence we apply the algorithm recursively on the (shorter) history $\ihistory2$ in which we have discarded the events involving the values $0$ and $1$. 
In the next step, we are facing the more complicated case, where the input history $\ihistory2$ has no extreme elements.
\smallskip 

Here, we will conceptually divide the history into a sequence of alternating segments, called $P$- and $D$-segments. A $P$-segment (P for \underline{p}opulated) is a maximal interval in which
there is at least one push operation which has returned, but the corresponding pop has not yet been called (hence the stack contains at least that element).  
In other words, it is the union of the I-segments of a set of values.
The I-segments of these values overlap, and the set is maximal in the sense that no other I-segment overlaps with them.
Intuitively, it is a window when the stack cannot be empty (hence the name {\it populated}), since there is at least one push which has returned  but the corresponding  pop
has not yet returned.
The $D$-segments  (D for \underline{d}eserted) are the complements the $P$-segments, i.e., it is a maximal interval, such that all push operations which have been called either (i) are  yet to return, or (ii) their pop operations have been called. Both (i) and (ii) can contribute to an empty stack.  
In other words, this is the window when the stack can be empty (no push is completed): it does not intersect with the I-segment of any value.
Thus a $D$-segment ends at the return of some push, when the stack becomes populated, beginning a $P$-segment. 

%

%
The $D$-segments of $\ihistory2$ in \cref{overview:algorithm1:fig} (in green)
are the intervals $\mkinterval{3}{6}$, $\mkinterval{12}{15}$, $\mkinterval{22}{23}$, $\mkinterval{27}{29}$ and $\mkinterval{31}{35}$. 
We  observe that the $P$- and $D$-segments build a sequence of intervals that are (i) consecutive, (ii) disjoint, (iii) alternate between $P$ and $D$, (iv) start with $D$, (v) end with $D$, and (vi) cover the entire history.
We divide the $D$-segments into two parts, namely the  (i)  {\it external}, and (ii) {\it internal} $D$-segments.
The external are the left-most and right-most segments, i.e., the interval $\mkinterval{3}{6}$ and $\mkinterval{31}{35}$.
The internal $D$-segments are $\mkinterval{12}{15}$, $\mkinterval{22}{23}$ and $\mkinterval{27}{29}$.
We identify one of the internal $D$-segments (the algorithm will identify the left-most internal $D$-segment), i.e., $\mkinterval{12}{15}$.
We consider all operations whose pushes return before (or just on the left border of)  the interval, i.e., latest at the point $12$.
We put all these operations (with values 2,3) in the left partition $L$.
Furthermore, we put all the other operations (with values 4,5,6,7,8) in the right partition $R$.
In \cref{overview:algorithm1:fig}, we redraw $\ihistory2$  showing the left partition (in blue) and the right partition (in red).
The crucial property of the partitioning is that we can assume, without losing precision, that the linearization points of all the operations in $L$ precede the linearization points of all the operations in $R$.
This means that we can now consider the linearization of two sub-histories, namely $\ihistory4$, containing the operations in $L$, and $\ihistory5$, containing the operations in $R$.
Now, the algorithm runs recursively on $\ihistory4$ and $\ihistory5$. %
For $\ihistory4$, it finds out that both values $2$ and $3$ are extreme.
Therefore, it eliminates them, and obtains the empty history $\emptyword$ that is trivially linearizable.
For $\ihistory5$, the algorithm detects that the value $8$ is extreme, and hence it removes it and calls itself on the resulting history $\ihistory6$.
Since there are no extreme values in the latter, it partitions $\ihistory6$ 
wrt the internal $D$-segment [22,23] 
into a left part and a right part, in  similar manner to the case of $\ihistory2$, obtaining $\ihistory7$ resp. $\ihistory8$.
The algorithm detects that all values in these two histories are extreme, and hence reduces each one of them to the empty history, concluding linearizability.

\begin{figure}
\includegraphics[scale=0.6]{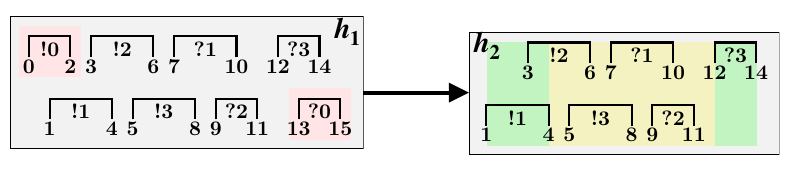}
\caption{The Stack algorithm concludes that $\ihistory1$ is unlinearizable.}
\label{overview:algorithm2:fig}
\end{figure}
In \cref{overview:algorithm2:fig}, we show how the algorithm concludes that the input history is unlinearizable.
In the first step it identifies the extreme value $0$ and removes it from $\ihistory1$ obtaining $\ihistory2$.
In the next step, it concludes that (i) $\ihistory2$ has no extreme values, and that (ii) $\ihistory2$ contains only two $D$-segments (with a $P$-segment in between). Then, it announces that $\ihistory2$, and hence also the input history $\ihistory1$, is unlinearizable.
As seen above, the algorithm proceeds to the next step if (a) there are extreme values, in which case,  it proceeds  by eliminating them, or (b)  there are three or more $D$-segments, in which case,  it proceeds   
by choosing the left most internal $D$-segment and partitioning around it.  Here, we have neither of these, and the algorithm concludes unlinearizability.
In general, a history is unlinearizable if (i) it has no extreme values, and (ii) it contains atmost two $D$-segments.
\begin{figure} \includegraphics[scale=0.55]{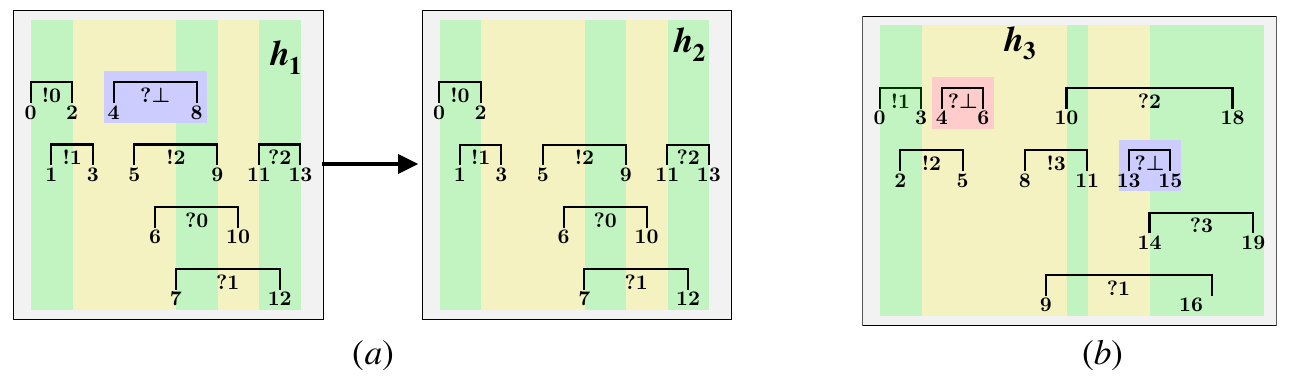}
\caption{Stack algorithm on Popempty : successful removal of $\popopof{\sbot}$ in (a), and unsuccessful in (b).}
\label{overview:popempty:algorithm2:fig}
\end{figure}


We consider the case where the history $\history$ contains an operation of the form $\popopof\sbot$.
In such a case, we systematically remove all occurrences of the operation in $\history$.
We consider two examples, (a) and (b) in  \cref{overview:popempty:algorithm2:fig}. 
There is one occurrence of the operation $\popopof\sbot$ in  \cref{overview:popempty:algorithm2:fig}(a). 
The only way to linearize $\ihistory1$ is to put the linearization point of $\popopof\sbot$ in one of the $D$-segments, since we require the stack to be empty to deal with 
$\popopof\sbot$. The algorithm goes through all the occurrences of the operation $\popopof\sbot$ (a single one in our case)  and ensures that it overlaps a $D$-segment.
The algorithm detects that the interval $\mkinterval48$ of the $\popopof\sbot$ operation intersects with the $D$-segment $\mkinterval79$.

It removes the operation, and calls itself recursively on the resulting history $\ihistory2$. In  \cref{overview:popempty:algorithm2:fig}(b), we have two occurrences of $\popopof\sbot$ in $h_3$. The one on the left occurs in $\mkinterval46$
which is inside the populated segment, and hence cannot succeed. The algorithm therefore concludes that $h_3$ is not linearizable. Notice that 
if we only had the second occurrence of $\popopof\sbot$ which occurs in [13,15], 
the algorithm would have detected its overlap with the $D$-segment [14,19], and 
removed it from $h_3$ as it did in \cref{overview:popempty:algorithm2:fig}(a), and 
continued calling recursively. Thus, to summarize, the algorithm  removes all occurrences of $\popopof\sbot$  unless it declares unlinearizability in the process.

\subsection{Queues}
\label{overview:queue:subsubsection}
\begin{figure} \includegraphics[scale=0.52]{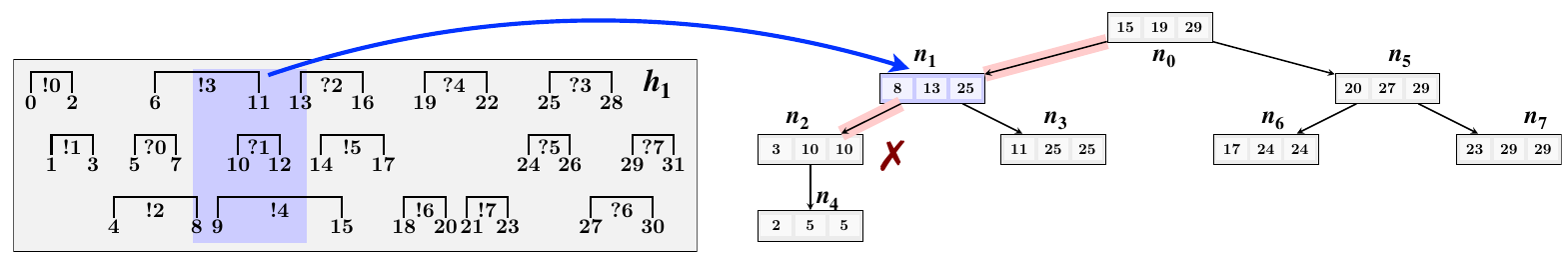}
\caption{Queue Algorithm : The blue arrow shows the node corresponding to the internal interval [8,13] of value 2. The red highlight on the tree edges show the branch chosen, culminating in  non-containment of the total interval [4,16] for value 2 with any other internal interval, as indicated by the red cross. 
}
\label{overview:queue:lin:algorithm:fig}
\end{figure}
\begin{figure} \includegraphics[scale=0.52]{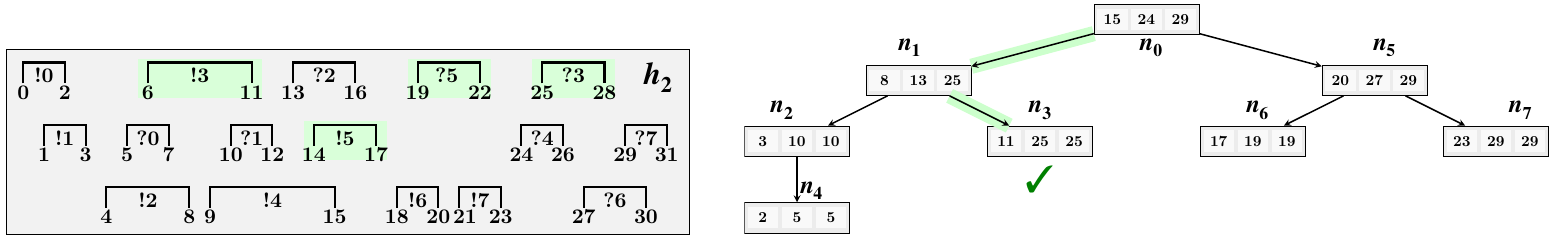}
\caption{Queue Algorithm : the total interval [14,22] of value 5 is contained in the internal interval [11,25] of value 3, leading to a critical pair (3,5) and unlinearizability. The green highlight on the tree edges show the branch chosen, culminating 
  in containment as indicated by the green tick. At $n_{1}$, we probe $n_{2}$ and see that its high key is too low to contain the interval $[11,25]$, which leads us to follow the edge to $n_{3}$.
}
\label{overview:queue:unlin:algorithm:fig}
\end{figure}

As we saw in \cref{overview:stack:subsubsection}, the stack algorithm relies on the definition of extreme values.
The algorithm for queues needs to replace  the last-in-first-out discipline, that holds for stacks, by the first-in-first-out discipline that holds for queues.
To reflect this difference, we  require that an extreme value $\val$ is one for which both the enqueue and dequeue operations of $\val$ are minimal, rather than being minimal resp.\ maximal.  
This results in a substantially different algorithm.
We will solve the queue problem in two steps.
First, we prove a {\it small model} property for the linearizability of queues. 
Second, we design a data structure, based on interval trees \cite{interval:trees}, that exploits the small model property to provide an ${\mathcal{O}}(n \log n)$ solution where $n$ is the number of operations in the input history. 

The small model property states that if a history $\history$ is unlinearizable wrt.\ $\queueadt$ then there is a {\it critical pair}, i.e., two values in $\history$ that are mutually unlinearizable wrt.\ $\queueadt$. A similar notion is present in earlier work \cite{AspectOrientedCONCUR13}. Here also, we use the notions
of $I$-segments (interval between the enqueue-return and dequeue-call time-stamps) and $T$-segments (interval between enqueue-call and dequeue-return time-stamps); these were defined for stacks in \cref{value:centric:subsubsection}. 
Notice that a pair is critical iff the $T$-segment of one of them lies entirely in the $I$-segment of the other.
Consider the history $\ihistory2$ in \cref{overview:queue:unlin:algorithm:fig} which is unlinearizable.
We can trace the reason for $\ihistory2$'s unlinearizability to the values $3$ and $5$ whose operations are shaded (in green) in the figure.
We observe that that 5's operations are ``sandwiched'' between 3's operations, hence violating the queue semantics.
On the other hand, the history $\ihistory1$ in \cref{overview:queue:lin:algorithm:fig} is linearizable and hence it does contain any critical pairs.
It is worth noting that the small model property does not hold for stacks.
For instance, the history $\ihistory1$ in \cref{overview:algorithm2:fig} is unlinearizable wrt $\stackadt$.
However, it does not contain critical pairs, i.e., there is no pair of values in $\ihistory1$ that mutually violate $\stackadt$. In fact, we can extend this construction to obtain the following lemma.
\begin{restatable}{lemma}{smallmodel}
  For any number $n \geq 2$, there is an unlinearizable stack history with $n$ values, for which any $n-1$-valued subhistory is linearizable.
\end{restatable}

To check linearizability we need to compare the $T$- and $I$-segments of all value pairs.
Doing the comparison naively will obviously result in a $\mathcal{O}(n^2)$ complexity.
To avoid the quadratic complexity, we propose a data structure that allows to perform the comparisons in $\mathcal{O}(n \log(n))$ time.
The data structure is a tree where the nodes represent the $I$-segments of the values in $\history$.
Crucially, we sort the nodes according to the left points of their $I$-segments, i.e., according to the time-stamps of the returns of their enqueue operations.
We build the tree to ensure the red-black property so that it allows search in logarithmic time.
Together with the interval, each node also stores the highest time-stamp that occurs in its sub-tree.
The tree $t_1$ in \cref{overview:queue:lin:algorithm:fig} corresponds to the history $\ihistory1$ in \cref{overview:queue:lin:algorithm:fig}.
Each node in $t_1$ represents  one value in $\ihistory1$.
A node is a labeled by a triple of numbers.
The left-most number is the left point of the value's $I$-segment; the number in the  middle is the right point of the value's $I$-segment; and the right-most number in the  largest time-stamp that occurs in the sub-tree of the node.
For instance, the node $\ignode1$ contains the $I$-segment of the value $2$, i.e., the interval $\mkinterval8{13}$, as its first and second elements.
Its third element is $25$ which is the highest time-stamp occurring in the sub-tree of $\ignode1$.

Once we have built the tree, we go through all the values again.
For each value, we check whether its $T$-segment lies within the $I$-interval of another value, i.e., within the interval of a node in the tree.
Let us consider the tree of \cref{overview:queue:lin:algorithm:fig}.
Assume that we are at the stage where we consider the value $2$.
The $T$-interval of $2$ is $\interval=\mkinterval4{16}$.
We start searching the tree from its root $\ignode0$.
We check whether the input interval $\interval$ is included in the interval of the root which is $\mkinterval{15}{19}$.
This is not the case, and hence we need to search the nodes in the rest of the tree.
The manner in which we have sorted the tree using the red-black property  implies that we need to search only one branch of the tree, by checking the current node (here, the node $\ignode0$) and its two immediate children, i.e., $\ignode1$ and $\ignode5$, we can decide whether we should go left or right.
The left-most number in $\ignode5$ is $20$.
Since we have sorted the tree according to their left-most numbers, the left points of  all the intervals in the  right sub-tree are larger than $20$.
Hence none of their intervals can contain $\interval$.
We conclude that we need to go left, so we move the search to the node $\ignode1$.
Again, we conclude that the input interval $\interval$ is not included in the interval of $\ignode1$ which is $\mkinterval8{13}$.
%
%
By a similar reasoning to above we conclude that we need to go left again, reaching $\ignode2$.
Here, we observe the right-most key in $\ignode2$ is $10$ which means that none of its sub-tree nodes can cover $\interval$.

In a similar manner, consider the tree in \cref{overview:queue:unlin:algorithm:fig}, 
and assume we are at the stage where we consider value 5. Then the total interval of 5 is [14,22].
Since the left-most point of the root $n_0$ is larger than 14, we search to the left. At $n_1$, 
the left-most value 8 is smaller than 14, but its high key is $10$, which is too low for any element in this subtree to contain the interval $[14,22]$. Thus, we move to the right subtree where we obtain the interval [11, 25] containing [14,22], and we conclude unlinearizability.


\section{Stacks}
\label{stacks:section}

In this section, we present the algorithm (\cref{stack:linearizable:algorithm}) for checking the linearizability of histories wrt. $\stackadt$, following the road-map of \cref{overview:stack:subsubsection}.
First, we define an algorithm to extract the value-centric view of the input history.
Then, we provide algorithms for calculating the $P$ and $D$ segments and use them to (i) find extreme values, (ii) partition the history into $P,D$ segments and (iii) check the validity of the placement of pop-empty operations. 
The proofs for all lemmas stated in this section are in the appendix.
\subsection{Completing the History}
\begin{definition}[Completion]
  The completion $\bar{h}$ of a history $h$ is obtained by adding concurrent pops at the end of the execution for each unmatched push.
\end{definition}
Intuitively, given a linearization of a history with unmatched pushes, the pushes will be in some order on top of the stack. If we were to start a pop for each such value, and then await their returns, we would end up with an empty stack. Formally, we have the following. 
\begin{restatable}[Completing an Execution]{lemma}{completioniff}
  A history $h$ is linearizable if and only if its completion $\bar{h}$ is linearizable.
\end{restatable}

Thus, from now on we consider only completed stack histories.
Similarly, any time the push and pop operations of a value overlap, we may ignore the value, since we may linearize them as $\pushopof{v}\app\popopof{v}$ at any point during the overlap. Thus, we also assume that the push and pop operations of any given value do not overlap. In the following, let $n$ be the number of operations in the given history. Each algorithm is simulated on our running examples in the Appendix.

\subsection{Calculating Value-Centric Views : \cref{optoval:algorithm}}
\begin{wrapfigure}{r}{0.5\textwidth}
\begin{algorithm}[H]
  \Input{A history $\history$}
  \Output{The set $\valvar$ of attributed values in $\history$}
  $\valvar\assigned\emptyset$\;
  \ForEach{$\op\in\history$}{
    $\val \assigned \valattrof\op$; 
    $\valvar\assigned\valvar\cup\set\val$\;
    $\callattrof{\val\cdot(\typeattrof\op)}\assigned \callattrof\op$\;
    $\retattrof{\val\cdot(\typeattrof\op)}\assigned \retattrof\op$\;
  }
  \Return $\valvar$
  \caption{$\optovalof\history$}
  \label{optoval:algorithm}
\end{algorithm}
\end{wrapfigure}
\cref{optoval:algorithm} computes the value-centric view of the input history $\history$ (see \cref{value:centric:subsubsection}). 
The algorithm calculates the set of attributed values by scanning all the operations occurring in $\history$ and storing data about their values
in the set $\valvar$.
More precisely, for each value $\val$, it calculates and stores its four time-stamps: push-call and push-return when it encounters the push operation on $\val$, and pop-call and pop-return when it encounters the pop operation on $\val$.
Complexity-wise, the outer loop is linear, and each access is logarithmic into a set structure, bringing the total of this algorithm to $\bigo(n~\log~n)$.
%

\subsection{Computing P-Segments : \cref{csegments:algorithm}}
\begin{algorithm}
  \Input{A set $\valvar$ of attributed values (of some history $\history$)}
  \Output{The set $\crowdedvar$ of $\history$}
  \tcp{Sort in a list  according to the order of push-return time-stamps} 
  $\listvar\assigned\sortalgof\valvar$
  \codespace
  $\crowdedvar\assigned\emptyset$
  \codespace
    $\currentvar\assigned [\pushretattrof{\listvar[0]},~ \popcallattrof{\listvar[0]}]$\;
  \For{$\ii\assigned1$ \KwTo $\sizeof\listvar-1$}{
    $\val \assigned \wposition\listvar\ii$\;
      \If{$\pushretattrof\val\leq\rightattrof\currentvar$}{
        $\jj\assigned\max((\popcallattrof\val),(\rightattrof\currentvar))$\;
        $\currentvar=\mkinterval{(\leftattrof\currentvar)}{\jj}$
      }
      \Else{
        $\crowdedvar\assigned\crowdedvar\cup\{\currentvar\}$\;
        $\currentvar\assigned\mkinterval{(\pushretattrof\val)}{(\popcallattrof\val)}$
      }
     }
     $\crowdedvar\assigned\crowdedvar\cup\{\currentvar\}$\;
  \Return $\crowdedvar$
  \caption{$\csegmentsalgof\valvar$}
  \label{csegments:algorithm}
\end{algorithm}
\cref{csegments:algorithm}  computes the $P$-segments.
The algorithm receives the value-centric view of the history, i.e., the set $\valvar$ of attributed values: the values that occur in the history, each together with its four time-stamps. 
The algorithm returns a set $\crowdedvar$ of intervals that are the $P$-segments of $\history$.
First, it sorts the set $\valvar$ in a list $\listvar$ according to the order in which their push operations return.
It scans the sorted list $\listvar$ and keeps track of the $P$-segment that is currently under construction, using a variable $\currentvar$.
The latter is initialized to the interval $[\pushretattrof{\val}, \popcallattrof{\val}]$ where 
$\val$ is the 
first value in the sorted list $\listvar$ (that is, $\pushretattrof{\val}$ is the 
earliest among all the push returns). 
If the I-segment of the following value overlaps with the $P$-segment under construction, we merge them into one and proceed.
Otherwise, we conclude that we are done with the current $P$-segment and move on to the next. The sorting is only done once, and we can keep this ordering across recursive steps of the main algorithm. This incurs in a preprocessing cost of $\bigo(n~\log~n)$. The loop is linear, and each step is constant as we store $\crowdedvar$ as an array with amortized constant time append.

\subsection{Computing D-Segments : \cref{esegments:algorithm}}

\begin{algorithm}
  \Input{A set $\crowdedvar$ of disjoint intervals representing the $P$-segments of a history $\history$}
  \Output{The set $\desertedvar$ of $D$-segments induced by $\crowdedvar$.}
  \tcp{Sort the set $\crowdedvar$ in a list according to left points of the intervals}
  $\min\assigned\min\setcomp{\pushcallattrof\op}{\op\in\history}$\;
  $max\assigned\max\setcomp{\popretattrof\op}{\op\in\history}$\;
  $\listvar\assigned\sortalgof\crowdedvar$
  \codespace
  $\desertedvar\assigned\mkinterval{\min}{(\leftattrof{\wposition\listvar0})}$\;
  \For{$\ii\assigned1$ \KwTo $\sizeof\listvar-1$}{
    $\desertedvar\assigned\desertedvar\cup\set{
      \mkinterval{(\rightattrof{\wposition\listvar{\ii-1}})}
                 {(\leftattrof{\wposition\listvar\ii})}}$
  }
  $\desertedvar\assigned\desertedvar\cup\set{
    \mkinterval{(\rightattrof{\wposition\listvar{\lengthof\listvar-1}})}{
               max}}$\;
  \Return\ $\desertedvar$
  \caption{$\dsegmentsalgof{\history,\crowdedvar}$}
  \label{esegments:algorithm}
\end{algorithm}
\cref{esegments:algorithm} inputs the set $\crowdedvar$ of $P$-segments and  returns the set $\desertedvar$ of $D$-segments. 
First, we sort the $P$-segments according to the left time-stamps (or, equivalently, the return time-stamps of pushes) of intervals. 
We build the set of $D$-segments by taking the right time-stamp of each $P$-segment and the left time-stamp of the next  $P$-segment. As with the previous algorithm, the list can be sorted ahead of time in a preprocessing step, and its order kept across recursive steps. The loop is linear, and each step is constant as we store $\desertedvar$ as an array with amortized constant time append.

\subsection{Checking Pop-Empty : \cref{pop:empty:algorithm}}
Handling $\popopof{\sbot}$-operations is done by scanning the history for such operations, and removing them if they intersect a D-segment. We show this is sufficient in the following lemma.
\begin{restatable}[Empty Correctness]{lemma}{emptycorr}
    \label{lem:opening_correctness}
    Let $h$ be a history with a $\popopof{\bot}$ operation $o$, i.e. a pop empty operation, and let $I_{o}$ denote the interval between the call and return of $o$. $h$ is linearizable if and only if there is a D-segment $\xi \in h$, such that $\xi \cap I_{o} \neq \emptyset$, and $h - o$ is linearizable.
\end{restatable}

 \begin{algorithm}
   \Input{A history $\history$, together with its set $\desertedvar$ of $D$-segments}
   \Output{Either conclude unlinearizability, or remove all occurrences of pop-empty}
   \ForEach{$\op\in\history$}{
     \If{$\eventattrof\op=\popopof\sbot$}{
       $\interval \assigned \intervalattrof\op$\;
         $\flagvar\assigned\false$\;
         \ForEach{$\binterval\in\desertedvar$}{
           \If{$\interval\intervalintersects\binterval$}
                       {
             $\flagvar\assigned\true$           
           }   
         }\tcp{$\intervalintersects$ represents a non-empty intersection between intervals}
         \If{$\neg\flagvar$}{
           \Exit{unlinearizable}
         }
         \Else{
           $\history\assigned\history-\set\op$
         }
       }
     }
      \Return{\history}
 \caption{$\popemptyof{\history,\desertedvar}$}
  \label{pop:empty:algorithm}
\end{algorithm}
  
The algorithm inputs a history $\history$ and its set $\desertedvar$ of $D$-segments.
It either exits declaring $\history$ unlinearizable or returns a history in which it has eliminated all occurrences of the $\popopof\sbot$ in $\history$.
To that end, it goes through the history operations one by one.
Each time it finds an operation $\op$ of the form $\popopof\sbot$, it scans through all the $D$-segments and checks whether at least one on them intersects with the interval of $\op$.
If this is the case, it discards $\op$ from the history.
Otherwise, it declares $\history$ to be unlinearizable.
The iteration is linear in the number of operations, and the intersection check in $\desertedvar$ is also linear.
The removal of an operation from $h$ takes $\log~n$. Thus, we have that the entire procedure for handling $\popopof{\sbot}$ is $\bigo(n (n + \log~n)) = \bigo(n^{2})$.

%

\subsection{Finding Extreme Values : \cref{extreme:algorithm}}
\begin{algorithm}
  \tcp{Sort the set $\desertedvar$ in a list according to left points of the intervals}
  $\listvar\assigned\sortalgof\desertedvar$
  \codespace
  $\extremevar\assigned\emptyset$\;
  \For{$\val\in\valvar$}{
    $\interval \assigned \mkinterval{(\pushcallattrof\val)}{(\pushretattrof\val)}$
    and
    $\binterval\assigned \mkinterval{(\popcallattrof\val)}{(\popretattrof\val)}$\;
    {
      \lIf{$\interval\intervalintersects\wposition\listvar0$
        and
        $\binterval\intervalintersects\wposition\listvar{\lengthof\listvar}$
      }{
      $\extremevar\assigned\extremevar\cup\set\val$
      }
    }
    \Return $\extremevar$
  }
  \caption{$\extremealgof\valvar\desertedvar$}
  \label{extreme:algorithm}
\end{algorithm}
Given the set $\valvar$ of attributed values and the set $\desertedvar$ of $D$-segments of a history $\history$, the algorithm finds the set $\extremevar$ of extreme values.
It first identifies the first and last $D$-segments.
Then it searches for values that occur in both of these and adds them to $\extremevar$. The first loop is linear in the number of values, and each check is constant time, thus the procedure of identifying extreme values is $\bigo(n)$.
\subsection{Partitioning : \cref{partition:algorithm}}
\begin{wrapfigure}{r}{0.5\textwidth}
  \begin{algorithm}[H]
  \ForEach{$\val\in\history$}{
    \If{$\pushretattrof\val \leq \leftattrof\interval$}{
      $\tmpvar\assigned\tmpvar\cup\set\val$
    }
  }
  \Return $\tuple{\tmpvar,\history-\tmpvar}$
  \caption{$\partitionalgof\history\interval$}
  \label{partition:algorithm}
  \end{algorithm}
\end{wrapfigure}
In \cref{partition:algorithm}, we take the first step to linearizability, by partitioning the history.  The algorithm 
takes parameters $h, \alpha$ for a history $h$ and an interval $\alpha$. 
For each operation $\op$ in the history, we check if the return point of its push 
is smaller than the left end point of $\alpha$, and if so, add it to $\tmpvar$. 
%
 The history is partitioned into two sub histories by considering 
on one hand,  all operations whose push return is before the left end of $\alpha$
and on the other hand, the remaining operations whose 
push return is at least at the left end of $\alpha$.
The outer loop is linear, and storing into $\tmpvar$ is amortized constant time, thus this algorithm is $\bigo(n)$.

\begin{restatable}[Partition Correctness]{lemma}{partitioning}
  \label{lem:sep_correctness}
  Given a partitioning of a history $h$ into $L, R$, we have that $h$ is linearizable if and only if $L$ and $R$ are linearizable.
\end{restatable}


\subsection{Main Algorithm for Stack: \cref{stack:linearizable:algorithm}}
\begin{wrapfigure}{r}{0.5\textwidth}
\begin{algorithm}[H]
  \lIf{$\history=\emptyword$}{
    \Return $\true$
  }
  $\valvar\assigned\optovalof\history$\;
  $\crowdedvar\assigned\csegmentsalgof\valvar$\;
  $\desertedvar\assigned\dsegmentsalgof{\history,\crowdedvar}$\;
  $\history\assigned\popemptyof{\history,\desertedvar}$\;
  $\extremevar\assigned\extremealgof\valvar\desertedvar$\;
  \If{$\extremevar\neq\emptyset$}{
    \Return $\linearizablealgof{\history-\extremevar}$
  }
  \Else{
        \lIf{$\sizeof\desertedvar\leq2$}{
      \Return $\false$
    }
    \Let{$\interval\assigned\wposition\desertedvar1$}{
      $\tuple{\ihistory1,\ihistory2}\assigned\partitionalgof\history\interval$\;
      \Return $\linearizablealgof{\ihistory1} \wedge \linearizablealgof{\ihistory2}$
    }
  }
  \caption{$\linearizablealgof\history$}
  \label{stack:linearizable:algorithm}
\end{algorithm}
\end{wrapfigure}
We now describe the main algorithm. If the history is empty, we conclude 
linearizability. Otherwise, using \cref{optoval:algorithm}, \cref{csegments:algorithm} and \cref{esegments:algorithm}, we populate respectively, $\valvar$, $\crowdedvar$
and $\desertedvar$.  Then we check if there are any $\popemptyof{\sbot}$ 
operations in the history, and remove them if possible, using \cref{pop:empty:algorithm}. Next, we check for extreme values 
in the history using \cref{extreme:algorithm}. In case there are any extreme values, 
the algorithm calls itself recursively on the history obtained by removing
the extreme values.
In case there are no extreme values, then the algorithm checks
for the number of $D$-segments in $\desertedvar$. In case there are only at most 2
$D$-segments, then the algorithm concludes non linearizability. Otherwise, the algorithm selects the first internal $D$-segment $\alpha$ and partitions
the history into $h_1, h_2$ using \cref{partition:algorithm}. 
On each partition, the algorithm is then called recursively.
The correctness of this procedure is covered by the following lemma.

\begin{restatable}[PushPop Characterization]{lemma}{characterization}
    \label{lem:characterization}
    A differentiated history $h$ without $\popopof\bot$ is linearizable if and only if one of the following conditions hold.
    \begin{enumerate}
      \item
            $h = \epsilon$
      \item
            There is a minimal push operation on a value $x$ in $h$, whose corresponding pop is maximal, and $h - \set{x}$ is linearizable, or
      \item
            $h$ can be partitioned into two nonempty and disjoint histories $L, R$, such that no operation in $R$ happens-before an operation in $L$, and each of $L$ and $R$ are linearizable.
    \end{enumerate}
  \end{restatable}

The first case of the lemma corresponds to the empty history. The second case corresponds to the removal of extreme values. The third case corresponds to partitioning around a $D$-segment. These are the three paths that can be taken by the algorithm.

%
We have the following results.
\begin{restatable}[Stack Algorithm Correctness]{theorem}{stackcorrect}
  The stack algorithm concludes linearizability if and only if the input history is linearizable.
\end{restatable}

\begin{restatable}[Stack Algorithm Complexity]{theorem}{stackcomplexity}
  The stack algorithm has time complexity $\bigo(n^{2})$ and space complexity $\bigo(n)$.
\end{restatable}

\section{Queues}
\label{queues:section}
This section presents the queue algorithm, following the ideas we presented in \cref{overview:queue:subsubsection}.
First, we introduce a data structure, called {\it Queue-Trees}, or {\it Q-Trees} for short, to define the algorithm.

\subsection{Queue Trees}
A {\it Q-Tree} is a (possibly empty) binary red-black tree where the set of nodes represents the values occurring in history.
A node $\gnode$ does not store the explicit value it represents (not needed in our algorithm) but instead stores five attributes, three of which are integers and two Q-Trees.
The attributes are defined as follows:
\begin{itemize}
\item
$\lsubattrof\gnode$  is the left sub-tree, and $\rsubattrof\gnode$ is the right sub-tree of $\gnode$.
\item
$\lkeyattrof\gnode$ is the left key, $\rkeyattrof\gnode$ is the right key, and $\hkeyattrof\gnode$ is the high key of $\gnode$.
The left and right keys are the left resp.\ right points of the value's $I$-interval.
The high key is the highest key occurring in the subtree with the node as root.
\end{itemize}
The nodes in the tree are sorted according to the left keys of their intervals.
We use $\emptyword$ to denote the empty Q-Tree.
%
\begin{algorithm}
  \Input{The root $\gnode$ of a Q-Tree, and an interval $\interval$}
  \Output{Is $\interval$ included in any node interval in the sub-tree of $\gnode$?}
  \lIf{$\gnode=\nullnode$}{\label{search:false}\Return $\false$}
  $l \assigned \lkeyattrof\gnode,\;\;\;r\assigned \rkeyattrof\gnode,\;\;\;h \assigned \hkeyattrof\gnode$\;
   $L \assigned \lsubattrof\gnode,\;\;\;R \assigned \rsubattrof\gnode, \; 
    i=\leftattrof{\alpha},\; j=\rightattrof{\alpha}$         \;
      \lIf{$l \leq i \wedge j \leq r$}{\label{search:local} \Return $\true$}
      \lIf{$(L \neq \nullnode) \wedge (l \leq i) \wedge (j \leq \hkeyattrof{L})$}{\label{search:true}
        \Return $\true$}
      \lIf{$(L \neq \nullnode)\wedge (l \leq i) \wedge (j > \hkeyattrof{L})$}{\label{search:right}
        $\searchalgof{R}\interval$}
       \lIf{$ i <  l$}{\label{search:left}
        $\searchalgof{L}\interval$}
  
\caption{$\searchalgof\gnode\interval$}
\label{queue:search:algorithm}
\end{algorithm}

The algorithm inputs the root $\gnode$ of a Q-Tree, and an interval $\interval$, and checks
whether there is a node in the tree whose interval contains $\interval$. The algorithm only searches up to one branch of the tree at any step. Since red-black trees are balanced, this equates to a complexity of $\bigo(\log~n)$.
Intuitively, if the tree is empty, we return false (\cref{search:false}).
Otherwise, we compare $\interval$ with the keys of $\gnode$ and its children.
 If the algorithm detects that $\interval$  is already contained in the interval of the root,  we trivially return true (\cref{search:local}).
In \cref{search:true}, we conclude that (i) the left key of  $\gnode$ is smaller than the left end-point of  $\interval$, and (ii) the highest key in the left sub-tree of $\gnode$ is larger than the right end-point of $\interval$.
We conclude that the answer is $\true$ in such a case.
The reason is twofold: (i)  the tree is sorted according to the left points of its intervals, and hence every node in the left sub-tree has a left-most point that is smaller than the left end-point of $\interval$; and (ii) the highest key in the left sub-tree is larger than the right end-point of $\interval$.
We observe that (i) and (ii) imply at least one interval in the left sub-tree of $\gnode$ contains $\interval$.
In \cref{search:right}, we go right since the right end-point of $\interval$ is larger than the highest key in the left sub-tree
of $\gnode$, and hence, $\interval$ cannot be contained in the intervals of the left sub-tree of $\gnode$.
Finally, we go left in \cref{search:left} since the left point of $\interval$ is smaller than the left key of $\gnode$.
Since the tree is sorted according to the nodes' left keys, all the nodes in the right sub-tree have 
left points that are higher than the left point of $\interval$, and hence they cannot contain $\interval$.

Formally, we show that the above algorithm works in the following lemma.
\begin{restatable}[Search Correctness]{lemma}{ctniff}
  \label{queue:lemma:ctniff}
  Let $\tree$ be a queue tree with root $\gnode$, and intervals $I$ (that is, $I-segments$)  and let $\alpha$ be an interval. We have
  \[
    \exists q \in I, \alpha \subseteq q \iff Search(\gnode)(\alpha)
  \]
  where $\alpha \subseteq q$ means $\alpha$ is contained in $q$.
\end{restatable}


\subsection{Main Algorithm}
\begin{wrapfigure}{r}{0.6\textwidth}
\begin{algorithm}[H]
  \Input{The (root of a) PQ-Tree $\gnode$}
  \Output{The completion of $\gnode$}
  \lIf{$\gnode=\nullnode$}{
    \Return $\nullnode$
  }
  $\ignode1\assigned\completionof{\lsubattrof\gnode}$\;
  $\ignode2\assigned\completionof{\rsubattrof\gnode}$\;
  $\hkeyattrof{\gnode}\assigned max(\hkeyattrof{\ignode1},\hkeyattrof{\ignode2},\rkeyattrof\gnode)$\;
\Return $\gnode$
\caption{$\completionof\gnode$}
\label{queue:completion:algorithm}
\end{algorithm}
\end{wrapfigure}
A {\it partial} Q-tree ($\buildpqtree$) is a a variant where the nodes have only the left and right keys, while the high key is missing. This is what is obtained from the standard algorithm for building a red-black tree with integer keys. We iterate through the $\buildpqtree$ to compute the high key for each node in linear time. We call this the completion of the $\buildpqtree$, and the algorithm is described in \cref{queue:completion:algorithm}.
%
%
We use a particular instance of the standard algorithm for building a red-black tree, $\buildpqtreeof\intervals$, which takes a set $\intervals$ of intervals and builds the red-black tree according to the left points of the intervals in $\intervals$.
The key of a node contains both the left point and the right point of each interval $\interval\in\intervals$ (although the sorting is carried only according to the left points).

\begin{algorithm}
  \Input{A history $\history$}
  \Output{$\history\models\queueadt$?}
  $\valvar\assigned\optovalof\history$\label{queue:alg:optoval}\;
  $\internalvar\assigned\emptyset$\;
  \lForEach{$\val\in\valvar$}{\label{queue:alg:internal}
    $\internalvar\assigned\internalvar\cup\set{\iattrof\val}$
  }
  \tcp{$\iattrof\val$ is the internal segment ($I$-segment) corresponding to value  $\val$}
  $\gnode\assigned\buildpqtreeof\internalvar$\label{queue:alg:RB}\;
  $\gnode\assigned\completionof\gnode$\;\label{complete}
  \For{$\val\in\valset$}{\label{queue:alg:for}
    \lIf{$\searchalgof\gnode{\tattrof\val}$}{
      \Return $\false$
    }\tcp{$\tattrof\val$ is the total segment ($T$-segment) corresponding to value  $\val$}
 
  }
  \Return $\true$
  \caption{$\linearizablealgof\history$}
  \label{queue:linearizable:algorithm}
\end{algorithm}
The algorithm inputs a history $\history$ and answers whether it is linearizable wrt. $\queueadt$.
In \cref{queue:alg:optoval}, the algorithm extracts the value-centric view from $\history$ similar to the case of stacks (\cref{stacks:section}).
In \cref{queue:alg:internal} it extracts the $I$-segments of the values and stores them in the set $\internalvar$.
In \cref{queue:alg:RB}, it builds the partial Q-Tree ($\buildpqtree$) 
described above and then completes (\cref{complete}) 
to obtain a Q-tree rooted at $\gnode$.
In \cref{queue:alg:for}, it goes through the values and checks if the $T$-segment of any one of them is included in some interval of the Q-tree.
In such a case, it declares the unlinearizability of the input history $\history$.
Else, it declares the linearizability of $\history$.
The algorithm performs $n$ (number of values) searches in the Q-tree. Each search has time complexity $\bigo(\log~n)$, which brings the total complexity of the algorithm to $\bigo(n~\log~n)$.
We rely on the following lemma for the correctness of the algorithm.
\begin{restatable}{lemma}{queuepair}
  \label{queue:lemma:pair}
  \[
    h\textrm{ linearizable } \iff \forall a,v, ~\tattrof{a} \nsubseteq \iattrof{v}
  \]
\end{restatable}


The lemma states that in order to check linearizability, it is sufficient to show that no $T$-segment is contained within any $I$-segment. This means that the checks performed in \cref{queue:linearizable:algorithm} are sufficient.

\begin{restatable}[Queue Algorithm Correctness]{theorem}{queuecorr}
  The queue algorithm concludes linearizability if and only if the input history is linearizable.
\end{restatable}

\begin{restatable}[Queue Algorithm Complexity]{theorem}{queuecomplexity}
  The queue algorithm has time complexity $\bigo(n~\log~n)$ and space complexity $\bigo(n)$.
\end{restatable}

\section{Sets and Multisets}
\subsection{Multisets}
The linearizability algorithm for multisets is less complicated than for the other data structures we cover.
We reuse the notation for stacks and queues, and state that for (multi)sets, $\pushopof{v}$ represents adding $v$, and $\popopof{v}$ represents removing $v$. Given a multiset history
$h$, a single value projection of $h$ is obtained from $h$ by projecting it down to any single value, and its operations from $h$. For a set history consisting of add and remove operations, we have the following.
\begin{restatable}[Multiset Projections]{lemma}{multisetpervalue}
  \label{lem:multiset_per_value}
  A multiset history $h$ is linearizable if and only if each of its single-value projections are linearizable.
\end{restatable}

For each of the resulting one-valued histories, we then have the following.

\begin{restatable}[Multiset Per-value]{lemma}{multisetsinglecount}
  \label{lem:multiset_single_count}
  A multiset history with a single value is linearizable if and only if at any time point, the number of returned removes does not exceed the number of called adds.
\end{restatable}

These two in combination yield a simple linear-time algorithm, outlined in \cref{alg:multiset}. Note that accessing the set of counters $C_{v}$ is done in amortized constant time using dynamic perfect hashing.

\begin{algorithm}
  \Input{A history $h$ as a sequence of calls and returns for events}
  $C_{v} \assigned (adds \assigned 0, rmvs \assigned 0)$\tcp*{Default counter values}
  \ForEach{$\event \in h$}{
    $v \assigned \valattrof{\event}$\;
    \lIf{$e = \callattrof{\pushopof{v}}$}{$C_{v}\cdot adds\assigned C_{v} \cdot adds+1$}
    \lIf{$e = \retattrof{\popopof{v}}$}{$C_{v} \cdot rmvs\assigned C_{v} \cdot rmvs+1$}
    \lIf{$rmvs > adds$}{
      \Return $\false$
    }
  }
  \Return $\true$
  \caption{MultisetLinearizable}
  \label{alg:multiset}
\end{algorithm}
\subsection{Sets}
\begin{wrapfigure}{r}{0.5\textwidth}
\begin{algorithm}[H]
  \Input{The sets $A_{v}$, $R_{v}$, $C_v$ and $S_{v}$, and a target state $q \in \set{\true, \false}$}
  \lIf{$S_{v} = q$}{\Return $\true$}
  \lIf{$q = \true$}{$\alpha = A_{v}$}
  \lElse{$\alpha = R_{v}$}
  $\alpha_{v}\cdot linearized \assigned \alpha_{v} \cdot linearized + 1$\;
  \lIf{$\alpha_{v} \cdot active < \alpha_{v} \cdot linearized$}{\Return $\false$}
  $C_{v} \assigned \emptyset$;
  $S_{v} \assigned q$\;
  \Return $\true$
  \caption{EnsureState}
  \label{alg:ensurestate}
\end{algorithm}
\end{wrapfigure}
We can do a similar count for sets, by also bounding the number of adds by the number of removes. Specifically, we require that, at each time point, the number of returned adds do not exceed the number of called removes by more than 1. From a history that satisfies this condition, it is easy to \emph{match up} the adds and removes to produce a linearization in which adds and removes alternate. On the other hand, if the condition is violated, it is easy to see that at some point, there must have been two adds in a row in any linearization.
We handle membership queries by tracking the number of active operations of each kind, as well as a state. A membership query is an operation $\gamma(v)$ that returns $\true$ if $v$ is present in the set, and $\false$ otherwise. We denote these two cases with $\gamma(v, \true)$ and $\gamma(v, \false)$, respectively. In essence, we handle each call and return in the order they appear in $h$. For calls, we increment $A_{v}\cdot active$ and $R_{v} \cdot active$ for adds and returns respectively. For calls to membership queries, we add the query to a set $C_{v}$. When we see a return, we check if the current state allows the returned operation to be linearized. If it does not, e.g. an operation $\pushopof{v}$ returned and the state $S_{v}$ is $\true$, we must first linearize a $\popopof{v}$-operation. We linearize the earliest returning such operation, and mark this by incrementing $R_{v}\cdot linearized$. We then linearize the returned add, and decrement the number of active adding operations. Whenever there is a change of state, we clear the set $C_{v}$, since each query operation can be linearized either before or after the change of state. When a membership query that has not been removed from $C_{v}$ returns, we check if the states match. If they do not, we perform the above procedure to switch the state and remove it.
\begin{algorithm}
  \SetKwFunction{FEns}{EnsureState}
  \Input{A history $h$ as a sequence of calls and returns for events}
  $A_{v} \assigned (active \assigned 0, linearized \assigned 0)$\;
  $R_{v} \assigned (active \assigned 0, linearized \assigned 0)$\tcp*{Default counter values for adds and removes}
  $C_{v} \assigned \emptyset$; $S_{v} \assigned \bot$\tcp*{Active membership queries and current state}
  \ForEach{$\event \in h$}{
    $v \assigned \valattrof{\event}$\;
    \lIf{$\event = \callattrof{\pushopof{v}}$}{$A_{v}\cdot active\assigned A_{v} \cdot active+1$}
    \lIf{$\event = \callattrof{\popopof{v}}$}{$R_{v} \cdot active\assigned R_{v} \cdot active+1$}
    \lIf{$\event = \callattrof{\gamma(v, x)} \wedge x \neq S_{v}$}{
      $C_{v}.\mathtt{insert}(o)$
    }
    \If{$\event = \retattrof{\pushopof{v}}$}{
      \If{$A_{v}\cdot linearized = 0$}{
        \lIf{$\neg$\FEns{$\false$}}{\Return $\false$}
        $C_{v} \assigned \emptyset$; $S_{v} \assigned \true$
      }
      \lElse{$A_{v} \cdot linearized \assigned A_{v} \cdot linearized - 1$}
      $A_{v} \cdot active \assigned A_{v} \cdot active-1$\;
    }
    \If{$\event = \retattrof{\popopof{v}}$}{
      \If{$R_{v}\cdot linearized = 0$}{
        \lIf{$\neg$\FEns{$\true$}}{\Return $\false$}
        $C_{v} \assigned \emptyset$; $S_{v} \assigned \false$
      }
      \lElse{$R_{v} \cdot linearized \assigned R_{v} \cdot linearized - 1$}
      $R_{v} \cdot active \assigned R_{v} \cdot active-1$
    }
    \If{$\event = \retattrof{\gamma(v, x)} \wedge o \in C_{v}$}{
      \lIf{$\neg$ \FEns{$x$}}{\Return $\false$}
      $C_{v}\cdot\mathtt{remove}(\eventattrof{e})$
    }
  }
  \Return $\true$
  \caption{SetLinearizable}
\label{alg:set}
\end{algorithm}

\paragraph{Failing Operations}
The common set definition includes that insertion and removal operations may fail. We can replace a failing $\pushopof{v}$ with an operation $\gamma(v, \true)$, since if it failed, it failed because the element was already present. We handle $\popopof{v}$ analoguously.

\begin{restatable}[Set Algorithm Correctness]{theorem}{setcorrectness}
  The set algorithm conclude linearizability if and only if the input history is linearizable.
\end{restatable}

\begin{restatable}[Set Algorithm Complexity]{theorem}{setcomplexity}
  The set algorithm has time complexity $\bigo(n)$ and space complexity $\bigo(n)$.
\end{restatable}

\section{Experiments}
\begin{figure}[ht!]
  \input{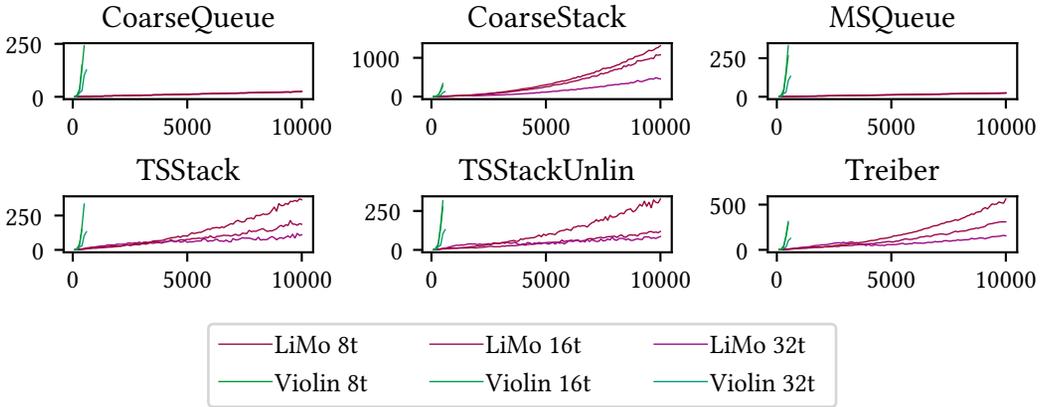}
  \caption{Experimental evaluation and comparison of histories of increasing size. The x-axis is the number of events, the y-axis is the slowdown in runtime relative to the smallest instance. 8t,16t, 32t represents 8,16,32 threads respectively.}
  \label{experiment:large}
\end{figure}
\begin{figure}[ht!]
  \input{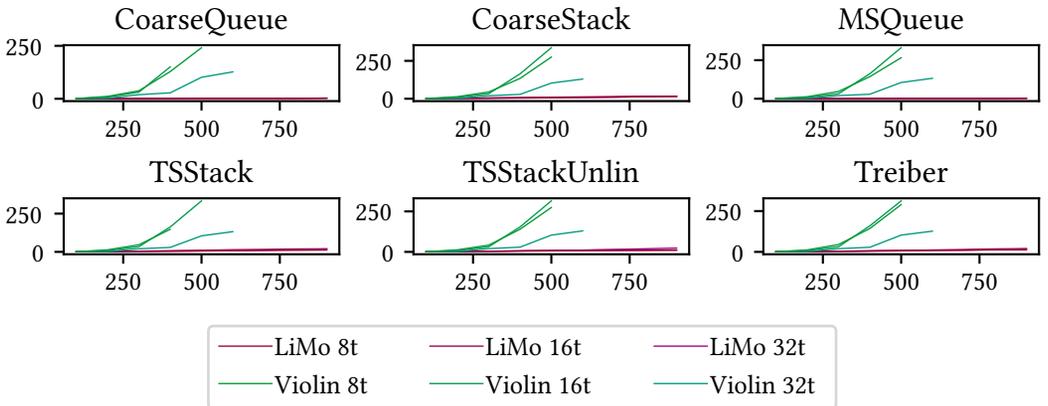}
  \caption{Experimental evaluation and comparison of smaller histories. The x-axis is the number of events, the y-axis is the slowdown in runtime relative to the smallest instance.}
  \label{experiment:small}
\end{figure}
In this section, we describe the implementation of our tool, $\limo$ (LInearizability MOnitor). We evaluate the scalability and efficiency of our algorithm by comparing it to previous work; namely $\violin$. As mentioned in the introduction, there is a bug in the  correctness proof. Although the underlying framework of Violin is not correct as pointed out in \cref{sec:related:issues}, we have not have not observed any bugs in the tool on the experiments we have conducted using 
$\violin$. $\violin$ is based on a saturation approach. Specifically, using a set of rules of the form
\[
  \pushopof{a} < \pushopof{b}~\wedge~\popopof{a} < \popopof{b} \implies \popopof{a}~<~\pushopof{b}
\]
together with transitivity, $\violin$ saturates the order $<$, which is initially the program order. If at any point there is a cycle in $<$, or equivalently $o < o$ for some operation $o$, $\violin$ concludes unlinearizability. Otherwise, $\violin$ concludes linearizability.

Due to the lack of other available tools, we do not make any other comparisons. We show that our algorithm is scalable and outperforms $\violin$, especially for larger histories.
\paragraph{Recording Executions}
We note that there is an inherent difficulty of obtaining accurate histories by recording executions without imposing artificial limits on concurrency. In order to test a given implementation, the call and return events must be recorded. The act of recording these events yield two problems. First, obtaining a total order of events requires synchronization of some sort. Second, the act of recording a call (or return) must occur strictly before (or after) the actual call and return of the operation, since some CPU instruction must be performed to acquire timing information. This means that each operation will inevitably become \emph{stretched}, which might introduce new possible linearizations due to the lost precision, possibly turning an unlinearizable history linearizable.

We have chosen to use a monotonous system clock, which guarantees that checking the current time will always yield monotonically increasing responses, possibly by sacrificing some accuracy in terms of real-world time. This is a good match for our use case, as we do not care about the specific times, merely the order between them. Using this clock we mark each event with a real-world time, which when collected and sorted yield a concurrent history.

\paragraph{Implementations}
Each of the two tools are tested on executions of four implementations of stacks. First, we have implemented a stack using coarse grained locks; i.e. \texttt{mutex} locks that cover each operation entirely. Second, we have implemented the well known \texttt{Treiber} stack \cite{treiber1986systems}. Third, we have implemented the time-stamped stack from \cite{DBLP:conf/popl/DoddsHK15}. Finally, we have produced a buggy implementation of the time stamped stack, that produces unlinearizable histories.

We also test the two tools on two queue implementations. One is an implementation using coarse grained locks, similar to the stack implementation of the same name. The other is the well known Michael-Scott queue \cite{DBLP:conf/podc/MichaelS96}.

\paragraph{Tested Histories}
We limit the execution time of both $\violin$ and $\limo$ to $5$ minutes.
Specifically, we evaluate on thread counts 8, 16 and 32. The number of operations varies from 100 to 10000, in increments of 100. For each size and thread count combination, we generate 10 histories. We only plot the results for which a given implementation solves all 10 histories of a given size without exceeding the timeout.
\paragraph{Fairness}
Due to programming language choices, $\limo$ has significantly lower run times, even when the amount of computation is roughly equivalent. It is several orders of magnitude faster in real-time. We have compensated for this by normalizing the times, dividing the runtime of an experiment on $k$ threads and $n$ operations with the runtime for the experiment on $k$ threads and $100$ operations, i.e. the first experiment with that thread count. This yields a slowdown plot, which is a more fair representation of the difference between the two \emph{algorithms}, rather than the two implementations.

\paragraph{Results}
Running the experiments on an Intel Xeon 8168 CPU running at 2.7GHz yields the results presented in \cref{experiment:large}. We see that $\violin$ quickly times out when increasing the number of operations. Since our tool has a lower initial running time, it can reach a much larger slowdown before timing out. Since $\violin$ times out quickly, we also plot the experiments when limited to 1000 operations in \cref{experiment:small}; this shows in a better manner, the scalability of $\violin$ compared to $\limo$. We see that $\limo$  greatly outperforms $\violin$, both for shorter and longer histories.

\section{Conclusions and Future Work}
We have presented a set of algorithms for monitoring libraries that implement concurrent data structures.
For stacks, queues, and (multi-)sets, the algorithm complexities are quadratic, log-linear, and linear, respectively.
The new upper bounds significantly improve on existing algorithms that are of at least cubic time complexity and have correctness issues.
Our tool implementation confirms the superiority of our algorithm both in efficiency and scalability.
A natural direction for future work is to consider other data structures, such as maps, priority queues, counters, and snapshots.
When monitoring a real system, it is often useful to have an algorithm that is \emph{online}. Our set algorithm is already online, and we believe one can extend the queue algorithm to be online. It is interesting to see whether one can produce an online stack algorithm while keeping the complexity quadratic.
It is also interesting to investigate whether our results can be carried over to general classes of $\adt$s, such as the collection types of \cite{DBLP:journals/pacmpl/EmmiE18}.
Designing general algorithms is a relevant but challenging task since, as we see in this paper, changing the $\adt$ results in substantially different algorithms.

\section{Data-Availability Statement}
Our tool, $\limo$, is available as an artifact \cite{grahn_limo_2025}, and is openly available on GitHub.
\newpage
\bibliographystyle{ACM-Reference-Format}
\bibliography{bibdatabase}
\newpage
\appendix
\newpage 

\section{Simulation of Stack Algorithms}
In this section, we simulate our algorithms on our running examples. 

\subsection{Calculating Value-Centric Views : \cref{optoval:algorithm}}
Let us run the algorithm on the history $\ihistory1$ of \cref{overview:histories:fig}. We begin with $\valvar$ as the emptyset. Scanning the history left to right, we encounter in order, the calls of $\pushopof{0}, \pushopof{1}$ at time-stamps 0,1 followed by
their returns at time-stamps 2,3. This results in adding values 0,1 to $\valvar$ and also updating $\pushcallattrof 0, \pushretattrof{0}, \pushcallattrof 1$ 
and $\pushretattrof{1}$ respectively to 0,2,1,3. Likewise, 
$\popcallattrof 0, \popretattrof{0}, \popcallattrof 1$ 
and $\popretattrof{1}$ are updated respectively to 5,7,4,6.

\subsection{Computing P-Segments : \cref{csegments:algorithm}}
Consider the history $\ihistory{2}$ in \cref{overview:algorithm1:fig}. First, sorting $\valvar$ yields 
the sorted list $\ell=(2,3,4,5,6,7,8)$ with $\ell[0]=2$ and 
$\currentvar=[6,9]$. Now, moving on with the for loop with $\ell[1]=3$,
we observe that $\pushretattrof{3}=8 \leq \rightattrof{\currentvar}=9$. Hence, we consider $\max(\popcallattrof{3},9)=\max(12,9)=12$, and update 
$\currentvar=[6,12]$. Then we continue with $\ell[2]=4$, for which 
$\pushretattrof{4}=15 > \rightattrof{\currentvar}=12$. 
Then we proceed to the else part where we add $[6,12]$ to $\crowdedvar$, update $\currentvar$ to $[15,20]$ and continue with $\ell[3]=5$. Here, $\pushretattrof{5}=16 < \rightattrof{\currentvar}=20$, and $\max(\popcallattrof{5}, 20)=\max(21,20)=21$. We update $\currentvar=[15,21]$, and proceed to $\ell[4]=6$.
Now, $\pushretattrof{6}=17 < \rightattrof{\currentvar}=21$, 
and $\max(\popcallattrof{6}, 21)=\max(22,21)=22$. We update $\currentvar=[15,22]$, and proceed with $\ell[5]=7$. Here, $\pushretattrof{7}=23 > \rightattrof{\currentvar}=22$. We proceed to the else part, where we add 
$[15,22]$ to $\crowdedvar$, update 
$\currentvar$ to $[23,27]$ and move to $\ell[6]=8$. We have $\pushretattrof{8}=29 > \rightattrof{\currentvar}=27$; hence  proceed to the else part and add 
$[23,27]$ to $\crowdedvar$, and update $\currentvar=[29,31]$. The loop ends and 
we update $\crowdedvar$ as the four intervals : [6,12], [15,22], [23,27] and [29,31].

\subsection{Computing D-Segments : \cref{esegments:algorithm}}
Let us continue with the example in \cref{overview:algorithm1:fig} for which 
we have computed the $P$ segments wrt the history $\ihistory{2}$. Now we run Algorithm \ref{esegments:algorithm} on the computed $\crowdedvar$. To begin, we have $\ell=([6,12], [15,22], [23,27])$. $i$=3, $j$=35, and we initialize $\desertedvar$ as $[3, 6]$.  Then we enter the for loop with $\ell[1]=[15,22]$, and 
add $[12,15]$ to $\desertedvar$. Likewise, for $\ell[2]=[23,27]$, we add 
$[22, 23]$ to $\desertedvar$. Finally, we add $[27, 29], [31,35]$  thereby obtaining $\desertedvar=\{[3,6],[12,15], [22,23]$, $[27,29], [31,35]\}$.

\subsection{Checking Pop-Empty : \cref{pop:empty:algorithm}}

As explained in the overview in \cref{overview:popempty:algorithm2:fig}(a),
we remove $\popopof\sbot$ which has its interval as $\alpha=[4,8]$ since it intersects ($\alpha \intervalintersects \beta$) with  $\beta=[7,9] \in \desertedvar$. 
  Likewise, in the case of \cref{overview:popempty:algorithm2:fig}(b), the occurrence of $\popopof\sbot$ has an interval $\alpha=[4,6]$  
  and $\neg([4,6] \intervalintersects \beta)$ for any $\beta \in \desertedvar=\{[0,3], [10,11], [14,19]\}$.

\subsection{Finding Extreme Values : \cref{extreme:algorithm}}

For example, consider the history $\ihistory{1}$ from \cref{overview:algorithm1:fig}.  
 We start with the list $\ell$ of sorted 
 $D$-segmentsfor $h_1$,  that is, $\{[0,6], [12,15],[22,23],[27,35]\}$. For the value 
0, we have the intervals $\alpha=[0,2], \beta=[30,34]$. Clearly, 
 $\alpha \intervalintersects [0,6]$ and 
$\beta \intervalintersects [27,35]$. Likewise, for the value 1,  
we have $\alpha=[1,4], \beta=[25,32]$ with the same observation. 
Also, none of the other values have a non empty intersection 
with $[0,6]$ and $[27,35]$.  
Thus we conclude that 0,1 are the extreme values. 

\subsection{Partitioning : \cref{partition:algorithm}}

For example, consider the example \cref{overview:algorithm1:fig} with $\alpha=[13,35]$. Now, we consider the values one by one : it can be seen that $ \pushretattrof{2}=6 < \leftattrof{\alpha}$; likewise,  	$ \pushretattrof{3}=8  < \leftattrof{\alpha}$, and no other value has this. Hence, the history is partitioned into two parts as seen in \cref{overview:algorithm1:fig}, the pink left part consisting 
of operations whose values are 2,3 and the blue right part consisting of operations on the remaining values. 
\subsection{Main Algorithm for Stack: \cref{stack:linearizable:algorithm}}

Let us consider the example in \cref{overview:algorithm1:fig}. We have already illustrated all the steps computing $\valvar$, $\crowdedvar$, $\desertedvar$, $\extremevar$ on this example. There is no occurrence of $\popemptyof{\sbot}$ in this instance.  The extreme values 0,1 are removed, and 
the resultant history  is partitioned wrt the first internal $D$-segment obtaining 
the histories $h_4, h_5$ in \cref{overview:algorithm1:fig}. The algorithm is recursively called on $h_4, h_5$, until we obtain $h_7, h_8$ on further partitioning where the algorithm concludes it is linearizable.

\section{Theory}
In this section we state and prove the correctness of our algorithms.
\subsection{Stack}
The proof of correctness is divided into two parts.
\smallskip

The first part deals with characterizing linearizability. The second part is to show that our algorithm correctly uses these characterizations to conclude linearizability if and only if there is a linearization.
In order to formally restate the stack specification, we formulate the following predicates on traces.
\begin{definition}[Matched and Empty]
  We say a history $h$ is \emph{matched} (denoted $\match(h)$) if for every push in $h$, there is a pop of the same value. Similarly, the predicate $\hempty(h)$ holds whenever there is some $\popopof{\bot}$ operation in $h$. Recall that
  $\popopof{\bot}$ represents attempting to pop from an empty stack.
\end{definition}
Let $\data(u)$ denote the set of values occuring in a history $u$. First, we state the rules for a stack linearization which allows for missing pop operations. We name these rules for easy reference.
\begin{align*}
  R_{\epsilon} : &~\epsilon \in \stackadt\\
  R_{PushPop} : &~u \cdot v \in Stack \wedge \data(u) \cap \data(v) = \emptyset \wedge \match(u)\\
              &~\Rightarrow \pushopof{x} \cdot u \cdot \popopof{x} \cdot v \in \stackadt\\
  R_{Push} : &~u \cdot v \in Stack \wedge \data(u) \cap \data(v) = \emptyset \wedge \neg \hempty(v)\\
              &~\Rightarrow u \cdot \pushopof{x} \cdot v \in \stackadt\\
  R_{PopEmpty} : &~u \cdot v \in Stack \wedge \match(u)\\
              &~\Rightarrow u \cdot \popopof{\bot} \cdot v \in \stackadt
\end{align*}

First, we show that the completion of a history preserves linearizability. This allows us to ignore $R_{Push}$ and $\match$
\completioniff*
\begin{proof}
  $(\Longrightarrow):$Assume $h$ is linearizable. Then, for every such linearization, there is some order of pops that removes the remaining values in an order opposite to the order in which they were pushed. Since all the relevant pops are concurrent, we may linearize them in this order, and obtain a linearization of the completion.
\smallskip

  $(\Longleftarrow):$ Assume $\bar{h}$ is linearizable with linearization $\ell$. Fix $\ell'$ as a sequentialization of $h$ that inherits the linearization points for each operation in $\bar{h}$. Then, $\ell'$ is a linearization of $h$, since each application of $R_{PushPop}$ may be replaced with $R_{Push}$.
\end{proof}

Now we assume we have a history for which $\match$ holds, and can thus safely ignore the rule $R_{Push}$.
An important property of linearizations is that under certain conditions, they can be concatenated into another linearization.
\begin{lemma}[Concatenation]
  \label{lem:concat}
  The concatenation of a linearization $\ell_{u}$ of $u$ and a linearization $\ell_{v}$ of $v$, is a linearization of $u \cdot v$.
\end{lemma}
\begin{proof}
  We will use induction on the size of $u$. 
  If it is empty, we are done. Assume it has size $n$. Consider the last applied rule for $u$. If it is $R_{PushPop}$, then we have that $u = \pushopof{x} \cdot u' \cdot v'$ for some $x, u', v'$. By induction we have that $v' \cdot v$ is a linearization, which in turn makes $\pushopof{x} \cdot u' \cdot \popopof{x} \cdot (v' \cdot v)$ a linearization of $u \cdot v$.
  The case for $R_{PushPop}$ is identical, and the case $R_{Push}$ is irrelevant since $u$ is matched.
\end{proof}

Now, we have shown that given a matched history, the above set of rules is equivalent to the rules presented in the paper.

\begin{definition}[Openings and Covers]
	In the following, we use the term Cover for $P$-segment and Opening for $D$-segments 
	interchangeably. 
\end{definition}

\begin{lemma}[Extreme Values]
  A value $a$ is extreme in a history $h$ if and only if its push starts before the first $P$-segment, and its pop ends after the last $P$-segment.
\end{lemma}
\begin{proof}
  $(\Longrightarrow):$ Assume $a$ is extreme, i.e. both minimal and maximal in $h$. Then, since it is minimal, it must necessarily start before any other push returns. Since a cover can only start via returns of a push, we conclude $\pushopof{a}$ must be started before the first $P$-segment of $h$.
  Similarly, since it is maximal, there is no pop that starts after the return of $\popopof{a}$. This implies there can not be any $P$-segment that ends after the return of $\popopof{a}$, and we are done.
\smallskip

  $(\Longleftarrow):$ Assume $\pushopof{a}$ starts before the first $P$-segment of $h$, and $\popopof{a}$ ends after the last $P$-segment of $h$. Then, there can be no push that returns before the call of $\pushopof{a}$, as such a push would have started a $P$-segment. This means $a$ is minimal.

  Similarly, there can be no pop that is called after the return of $\popopof{a}$, since then there would have been a $P$-segment that covered the entirety of $\popopof{a}$. Thus, there can be no operation happening strictly after $\popopof{a}$, and it is thus maximal.
\end{proof}

Next, the partitioning of a history by an $D$-segment is equivalent to its projection onto two disjoint sets with some additional properties. To use this, we will formulate the following definition.

\begin{definition}[Projection]
  Projecting a history $h$ onto a set of values $V$, by keeping only the values in $h$ that lie in $V$, forms a sub history $\proj{h}{V}$ of $h$.
\end{definition}

\begin{lemma}[Projection Preserves Linearizability]
  \label{lem:proj_lin}
  A history $h$ is linearizable if and only if all of its projections are linearizable.
\end{lemma}
\begin{proof}
  $(\Longleftarrow):$ The projection $\proj{h}{\data(h)} = h$, is by assumption linearizable, and we are done.
\smallskip
  $(\Longrightarrow):$ Assume $h$ is linearizable and fix an arbitrary set $S$. We will perform structural induction on the stack rules.
  Consider a linearization of $h$, and consider the last rule applied to construct it.
  If it is $R_{PushPop}$, we have two cases. If the witness for this rule, (i.e. the value $x$) is in $S$,
 then, by induction, we can obtain linearizations $\ell_{u}, \ell_{v}$ of $\proj{u}{S}$ and $\proj{v}{S}$, respectively.  We then have a linearization
 $\pushopof{x} \cdot \ell_{u} \popopof{x} \cdot \ell_{v}$.
  Otherwise, if $x \notin S$, we again note that we can obtain two linearizations $\ell_{u}$ and $\ell_{v}$ of $\proj{u}{S}$ and $\proj{v}{S}$, respectively, by induction. From \cref{lem:concat}, we have that this is a linearization of $\proj{h}{S}$.

  Otherwise, the last rule applied is $R_{PopEmpty}$. Once again we have two cases. Either $\bot \in S$, at which point we may immediately construct the linearization of $\proj{h}{S}$, similar to the previous case. Otherwise, we conclude that by concatenating the two linearizations $\ell_{u}, \ell_{v}$ of $\proj{u}{S}$ and $\proj{v}{S}$, respectively, we obtain  a linearization of $\proj{h}{S}$ (\cref{lem:concat}).
\end{proof}

This characterization is equivalent to that of \cite{JAD18}. In particular, we refer to two of the characterizations, relating to $R_{PopEmpty}$ and $R_{PushPop}$.

\characterization*
  \begin{proof}
    This lemma is equivalent to Lemma 16 in \cite{JAD18}.
    \smallskip
    $(\Longrightarrow):$ Assume there is a linearization $\ell$ of $h$, and consider the last inductive rule applied to construct this linearization. We note that it is either $R_{\epsilon}$ or $R_{PushPop}$, since we assume $\match$ and that $h$ does not have a $\popopof{\bot}$-event.
    If the last rule applied is $R_{\epsilon}$, then we have that $h = \epsilon$, and we are done.
    If it is $R_{PushPop}$, i.e. $\ell = \pushopof{x} \cdot u \cdot \popopof{x} \cdot v$ for some $x$, then we have two cases.
    If $v = \emptyset$, then we have that $x$ is an extreme value, and we are done.
    If $v \neq \emptyset$, then $L = \pushopof{x} \cdot u \cdot \popopof{x}$ and $R = v$ is a separation, and we are done.
\smallskip
    $(\Longleftarrow):$ We have three cases. If $h = \emptyset$, then $\ell = \epsilon$ is a valid linearization. If there is a push operation on a minimal $x$ with maximal pop, and $h - \set{x}$ has a linearization $\ell$, then $\pushopof{x} \cdot \ell \cdot \popopof{x} \cdot \epsilon$ is a linearization of $h$. Otherwise, if there is a separation of $L, R$ into disjoint sub histories, with linearizations $\ell_{L}$ and $\ell_{R}$ respectively, we conclude that $\ell_{L} \cdot \ell_{R}$ is a linearization of $h$, since $L, R$ are disjoint and $\match$.
\end{proof}

\partitioning*
\begin{proof}
  $(\Longrightarrow):$
  Assume for contrapositive that either $L$ or $R$ is not linearizable. Then, there is a projection of $h$ onto the values of $L$ or $R$ that is not linearizable. Thus, from \cref{lem:proj_lin}, $h$ is also not linearizable.
  \smallskip
  $(\Longleftarrow):$ The concatenation of the linearizations for the $L$ and $R$ sub histories form a linearization for $h$.
\end{proof}

\begin{lemma}[Nonempty Partition]
    \label{lem:sep_nonempty}
    Separating around an internal $D$-segment yields two sub histories $L$ and $R$ that are each nonempty.
\end{lemma}
\begin{proof}
  Let $h$ be a history with an internal $D$-segment. Since the $D$-segment is inner, there must have been some $P$-segment before it. A $P$-segment is started by the return of a push operation. Thus, there is some push returning before the $D$-segment, and the set $L$ must be nonempty. Similarly, there needs to be some $P$-segment after it, which means some push needs to return after the $D$-segment. This value will be put in $R$. Thus, each of $L$ and $R$ are nonempty.
\end{proof}

\emptycorr*
\begin{proof}
  $(\Longrightarrow):$
  Assume $h$ is linearizable. Then, there must have been a point during the execution of $o$ that the stack was empty, and this must have been during a $D$-segment. Moreover, projection preserves linearizability, and we have that $h-o$ is linearizable.
  \smallskip
  $(\Longleftarrow):$
  Let $\xi$ be the opening, and assume $h - o$ is linearizable. Then, there must be some point during $\xi$ that the stack is empty. At this point, we can choose to linearize $o$, using $R_{PopEmpty}$, and we are done.
\end{proof}

Moreover, we state that the existence of inner openings coincide with separability.
\begin{lemma}[Inner Opening Separability]
  \label{lem:inner_sep}
  Let $h$ be a history. There is a separation of $h$ into nonempty $L, R$ if and only if there is an inner opening (inner $D$-segment) in $h$.
\end{lemma}
\begin{proof}
  $(\Rightarrow):$ Let $L, R$ be a nonempty separation of $h$.
  Let $\pushopof{r}$ be the earliest returning push in $R$, and let $\popopof{l}$ be the latest called pop in in $L$.
  However, we know from the definition of a separation that $\pushopof{r} \not < \popopof{l}$. In other words, we have
  $\callattrof{\popopof{l}} < \retattrof{\pushopof{r}}$. We claim the interval defined by these two points is an opening.

  In order to show this claim, we will show that for any $x$, its $P$-segment cannot intersect the interval $\interval{\callattrof{\popopof{l}}, \retattrof{\pushopof{r}}}$.

  If $x \in L$, we have that $\callattrof{\popopof{x}} < \callattrof{\popopof{l}}$ from the assumption that $l$ has the latest called pop in $L$, thus we know the $P$-segment of $x$ strictly preceeds the opening, and is thus not intersecting it.

  Similarly, if $x \in R$,  we have $\retattrof{\pushopof{x}} > \retattrof{\pushopof{r}}$, from the assumption that $r$ is the earliest returning push in $R$, thus we know the $P$-segment of $x$ strictly succeeds the opening, and is thus not intersecting it.

  Thus, we conclude it is an opening. Moreover, it is an inner opening since there is a $P$-segment for $l$ (and for $r$) that preceeds (resp. succeeds) it.
  $(\Leftarrow):$ Assume there is an inner opening. From \cref{lem:sep_nonempty} we obtain a separation of $h$ into nonempty $L, R$.
\end{proof}

We also state the characterization for $R_{PopEmpty}$.

\begin{lemma}[PopEmpty Characterization]
  \label{lem:pemp}
  A history containing a $\popopof{-}$ operation is linearizable if and only if the operation is overlapping with some opening of $h$, and each of $L$ and $R$ obtained by separation over this opening is linearizable.
\end{lemma}
\begin{proof}
  $(\Longrightarrow):$ Assume a history $h$ with a $\popopof{-}$ operation $o$ is lineariable. Then, the stack must be empty at the linearization point of $o$. This means, there has to be an opening intersecting the linearization point of $o$. Separating over this opening yields two subhistories, $L$ and $R$, with $L$ containing all values occurring before $o$ in the linearization, and $R$ containing all values occurring after $o$ in the linearization. Moreover, by \cref{lem:proj_lin}, each of $L$ and $R$ are linearizable.
\smallskip

  $(\Longleftarrow):$ Let $\ell_{L}, \ell_{R}$ be linearizations of separations around an opening of $h$, that intersects $o$. We have that, since $L$ is matched, $\ell_{L} \cdot o \cdot \ell_{R}$ is a linearization of $h$, by \cref{lem:concat}.
\end{proof}

Now, we may finally state the main theorems.

\stackcorrect*
\begin{proof}
  $(\Rightarrow)$
  Assume the algorithm concludes linearizability. We will construct a linearization, with induction on the length of the history. In the base case, $h = \epsilon$ and $|h|=0$,  which is correctly identified as linearizable.
  If $|h| = n$, we have some cases.
  In the first case, $\hempty{h}$.
  In this case, the algorithm concludes linearizability if it can separate around some opening that intersects the $\popopof{\bot}$ operation, and conclude linearizability for both $L$ and $R$ contained this way. By induction, these conclusions guarantee the existence of linearizations $\ell_{L}, \ell_{R}$.
  Then, $\ell_{L} \cdot \popopof{-} \cdot \ell_{R}$ is a linearization of $h$.

  Thes second case is when $\neg \hempty{h}$.
  If the algorithm concludes linearizability, it does so because either there is some $x$ that is both minimal and maximal, and $h - \{x\}$ concludes linearizability, or there is a separation into $L$ and $R$, that the algorithm concludes linearizability for. By the induction hypothesis, each of these sub histories are of size less than $n$, so we conclude they are in fact linearizable. In the first case, $\pushopof{x} \cdot \ell_{h-x} \cdot \popopof{x}$ is a linearization of $h$. In the second case, $\ell_{L} \cdot \ell_{R}$ is a linearization of $h$, and we are done.

  ($\Leftarrow$)
  Assume for contrapositive the algorithm concludes that a history $h$ is not linearizable. Then, we locate the smallest sub history $h'$ where this is concluded.
  If it is concluded for a history without $\popopof{\bot}$-events, then there is neither an extreme value, nor an internal $D$-segment, which by \cref{lem:inner_sep} means there is no separation of $h'$ into $L, R$. From \cref{lem:characterization}, we know this implies that $h'$ is not linearizable, which by \cref{lem:proj_lin} implies $h$ is not linearizable.

  Otherwise, we have a $\popopof{\bot}$-event. Then, there is no opening $o$ that overlaps a given
  $\popopof{\bot}$ operation. From \cref{lem:pemp}, we know $h'$ is not linearizable, which by \cref{lem:proj_lin} means that $h$ is also not linearizable.
\end{proof}

\stackcomplexity*
\begin{proof}
  We start with time complexity. The pop-empty procedure is a one time check, incurring $\bigo(n^{2})$. Likewise, all steps involving sorting (of the sets $\crowdedvar, \desertedvar$, $\valvar$) is done ahead of recursion. Each call to the main algorithm involves computing the $\crowdedvar, \desertedvar$ which are already maintained in a sorted fashion. The total complexity of all steps taken before each recursive step is $\bigo(n)$. The history is then divided into one or two subhistories. In the worst case, the size of one of these subhistories is $n-1$, at which point the total complexity of the algorithm is $\bigo(\sum_{1}^{n} n) = \bigo(n^{2})$.

  Next, for space complexity, we keep the following in memory;
  \begin{itemize}
    \item
          For each operation, its relevant attributes, $\bigo(n)$.
    \item
          $\crowdedvar, \bigo(n)$
    \item
          $\desertedvar, \bigo(n)$
    \item
          $\valvar, \bigo(n)$
  \end{itemize}
  Which in total is $\bigo(n)$.
\end{proof}

Additionally, we present the proof of absence of a small model property for stacks.
\smallmodel*
\begin{proof}
  For $n = 2$, we have the trivial, sequential history $\pushopof{0} \prec \pushopof{1} \prec \popopof{0} \prec \popopof{1}$.

  Let $n \geq 3$ be a natural number, and let $\data = \set{1, \dots, n}$.
  We will construct a history with values $\data$, that is unlinearizable, but removing any one value from $\data$ yields a linearizable history.
  Consider the following history.\\
  \includegraphics[width=0.5\textwidth]{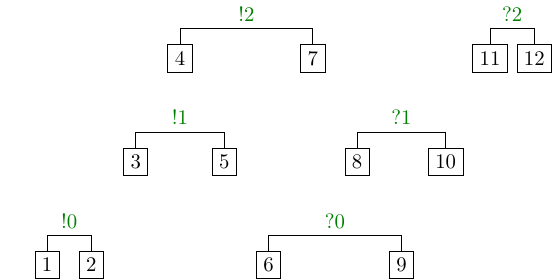}\\
  Notice that the $I$-segments of each value overlap with the \emph{next} value. Additionally, only $0$ is minimal and $2$ is maximal. This means that there is only one P-segment, and no value is both minimal and maximal, so the history is unlinearizable.

  In the general case, where we have $n$ values, we let the first value, $1$, have a minimal push. Following the minimal push, we have a call to each other push. Then, we stagger the returns of pushes and calls to pops, such that the I-segment of value $i$ overlaps only with the I-segments of value $i-1$ and $i+1$. We do this for each $i < n$. Finally, we have the returns to each pop operation, at which point we append the call and return of $\popopof{n}$.

  It is clearly unlinearizable, as there is no extreme value, and only one $P$-segment.

  To show that removing any one value gives a linearizable history, we first consider removal of one of $1$ and $n$. This makes the other an extreme value, at which point we can linearize it, and the remainder of the values are each both minimal and maximal, and we are done.
  The other case is when we remove some value $1 < i < n$. In such a case, we will create an inner P-segment. This gives a separation which places the value $1$ in L and the value $n$ in R. In L, $1$ will be extreme, so we can remove it, and the rest of the values in L will all be extreme. Similarly, in R, $n$ will be extreme, and removing it yields a history in which each other value is extreme.
\end{proof}

\subsection{Queues}
\ctniff*
\begin{proof}
  We note that both directions are trivial for $|\tree| \leq 1$. Thus, we perform induction, assuming the theorem holds for $|\tree| < n$.\\\\
  $(\Rightarrow):$ Assume there is such a $q$. We will show that the algorithm returns $\T$. We perform structural induction on $\tree$. We know $\tree \neq \emptyset$ since $q \in I$. We have three cases for where in the tree $q$ is located.
  \begin{enumerate}
    \item $\tree = q$, i.e. $q$ is the root interval of $\tree$. Then, we trivially have $q.l \leq L \wedge R \leq q.r$, and we are done.
    \item $q \in \lsubattrof\tree$. Then we know that $R \leq \lsubattrof{\tree}.k$.
          Thus we are in one of two cases.
          \begin{enumerate}
            \item $i \leq L$: We return $\T$.
            \item $L < i$: We recurse into $\ctn{\lsubattrof{\tree}, \intv{L, R}}$. By induction, this returns $\T$.
          \end{enumerate}
    \item $q \in \rsubattrof{\tree}$. We note that, if there is some $q' \in \lsubattrof{\tree}$ that contains $\intv{L, R}$, we have already proven that the algorithm produces a correct result. Thus, we consider the case when $q \in \rsubattrof{\tree}$, and no such $q'$ exists in $\lsubattrof{\tree}$.
          Then, the only cases that remain are
          \begin{enumerate}
            \item $L \leq i$, which cannot happen since then we would have $L \leq i \leq q.i \leq L$.
            \item $i \leq L \wedge R > \lsubattrof{\tree}.k$, at which point we return the recursive call $\ctn{\rsubattrof{\tree}, \intv{L, R}}$, which by induction returns $\T$.
          \end{enumerate}
  \end{enumerate}
  $(\Leftarrow):$ Assume $\ctn{\tree, \intv{L, R}} = \T$, and we wish to find a $q \in I$ such that $\intv{L, R} \subseteq q$. We know $\T \neq \emptyset$, so we have $\tree.root = (i, j, k)$. We split into cases.
  \begin{enumerate}
    \item $i \leq L \wedge R \leq j$, then the root value is such a $q$.
    \item $i \leq L \wedge R \leq \lsubattrof{\tree}.k$, then some interval in $\lsubattrof{\tree}$ has the form $(w, k, k)$, the value of which is a $q$ that satisfies the requirement, since $w \leq i \leq L \wedge R \leq k$.
    \item $i \leq L \wedge R > \lsubattrof{\tree}.k$. Then we recurse into $\ctn{\rsubattrof{\tree}, \intv{L, R}} = \T$, which by induction yields such a $q$.
    \item $L < i$. We recursively call $\ctn{\lsubattrof{\tree}, \intv{L, R}} = \T$, and by induction we have such a $q$.
  \end{enumerate}
\end{proof}

\subsubsection{Small Model Property}
\begin{lemma}
  If $h$ is a queue history, then $h$ is linearizable if and only if $h$ has an extreme value $x$, and $h - x$ is linearizable.
\end{lemma}
\begin{proof}
  $(\Leftarrow):$ If $x$ is an extreme value, then we know $deq(x)$ can be linearized before any other $deq$, and $enq(x)$ can be linearized before any other $enq$. Pick a linearization $\ell$ of $h - x$. It is of the form
  $enq^{*} \cdot \ell'$ for some $\ell'$. We state that $enq(x) \cdot enq^{*}\cdot deq(x) \cdot \ell'$ is a linearization of $h$.\\\\
  $(\Rightarrow):$ Assume $h$ linearizable. Then, by definition, the first operation must be an $enq(x)$ for some $x$. Moreover, the first deq must be on $x$, thus $enq(x)$ is a minimal enq, and $deq(x)$ is a minimal deq.
\end{proof}
\queuepair*
\begin{proof}
  $(\Rightarrow):$ Trivial, since otherwise, for the pair $a, v$ with $\tattrof{a} \subseteq \iattrof{v}$, we have the pattern $enq(v) \cdot enq(a) \cdot deq(a) \cdot deq(v)$, which is a violation. Since linearizability is preserved by projection, $h$ cannot be linearizable.
  $(\Leftarrow):$ We will perform induction on the size of $h$. The base case, $|h| \leq 2$ trivially hold. Assume $|h| > 2$.

  Let $a$ be the value of the minimal $enq$ with earliest called $deq$. If $deq(a)$ is minimal, we have that $a$ is extreme. Then, by the above lemma, we have $h$ linearizable if and only if $h - a$ is linearizable. By induction, we have that $h-a$ is linearizable.

  Otherwise, we have that there is some $q$ for which $deq(q) < deq(a)$. $q$ cannot be minimal, since this contradicts our choice of $a$. Thus, there must be some minimal $w$ for which $enq(w) < enq(q)$. However, since $a$ has the earliest called deq of all minimal enq, we must necessarily have $deq(q) < deq(w)$, which yields
  $enq(w) < enq(q) < deq(q) < deq(w)$, and we have $q \in cover(w)$, contradicting our initial assumption.
\end{proof}
\queuecorr*
\begin{proof}
  Follows from \cref{queue:lemma:ctniff} and \cref{queue:lemma:pair}.
\end{proof}

\queuecomplexity*
\begin{proof}
  We first construct the tree by adding $n$ elements. Each insertion is $\bigo(\log~n)$ since it is a red-black tree. Second, we compute the high key, which we can do in linear time by iterating over the nodes bottom up. Third, we check membership.

  Since the queue tree is a balanced tree, it is sufficient to show that the search algorithm only searches one branch, which is obvious from its definition.

  For space complexity, we note that we need only keep the tree and the T-segments in memory, each of which is $\bigo(n)$.
\end{proof}

\subsection{Multisets}
\multisetpervalue*
\begin{proof}
  Since we only allow the operations $\pushopof{v}$ and $\popopof{v}$, this holds trivially; the presence or absence of a value $x$ does not impact how the set should behave for a value $v \neq x$.
\end{proof}

\multisetsinglecount*
\begin{proof}
  $(\Longrightarrow)$: By contrapositivity assume that the number of returned removes exceeds the number of called adds. Then, at this point in time, there have been more remove operations than add operations, which trivially does not satisfy multiset.
  $(\Longleftarrow)$: Assume the the number of returned removes does not exceed the number of called adds. Define a linearization of $h$ by having each add operation be linearized at its call time, and each remove operation be linearized at its return time. For this linearization not to satisfy multiset, there must at some time point have been more removes than returns. However, our choice of linearization points together with our assumption prevents this.
\end{proof}

\setcorrectness*
\begin{proof}
  $(\Rightarrow):$ Assume the algorithm concludes linearizability. The algorithm implicitly produces a trace $\tau$. It suffices to show that $\tau$ is a linearization.
  To show this, we assume the contrary. In other words; assume that at some point in some single-valued projection of $\tau$, we have one of the following cases.
  \begin{enumerate}
    \item $\pushopof{v}_{2}$ happens after another $\pushopof{v}_{1}$, without a $\popopof{v}$ in between.
    \item $\popopof{v}$ happens after another $\pushopof{v}$, without a $\pushopof{v}$ in between.
    \item $\popopof{v}$ happens before the first $\pushopof{v}$, or
          The algorithm never linearizes a pop operation when the state is $\false$, or a push operation when the state is $\true$, so neither of these cases can happen.

    \item A $\gamma$-operation is linearized in the wrong state.

          This cannot be the case, since either it was linearized in EnsureState, at which point the states flipped and we chose to linearize it at the right position, or, it was linearized in its call or its return.

          If it was linearized in its call, the state was already correct and we are done. Similarly, if it was linearized in its return, it was again either linearized in EnsureState, or the state was correct.
  \end{enumerate}
\end{proof}

\setcomplexity*
\begin{proof}
  We start with time complexity. The algorithm iterates over $n$ events. Accessing the correct counters is done in amortized constant time using dynamic perfect hashing. Thus, we only need to show that each event handling procedure is constant time.

  The calls to $\pushopof{v}$ and $\popopof{v}$ are clearly linear, as it is only an increment operation.
  Similarly, the call to a $\gamma$-operation is constant time due to dynamic perfect hashing, and the return of it features a removal, which is also amortized constant time.

  Each other statement is either a call to EnsureState, or an assignement statement. Thus, the only thing remaining is to show that EnsureState is constant time.
  This is easy to see, since every statement is an assignment or a comparison.

  We proceed to show the space complexity. The algorithm stores, in total, 3 counters for each value, so $\bigo(n)$. The total combined size of all sets $C_{v}$ is at most equal to the number of operations, when every operation is a $\gamma$-operation. Thus the algorithm has linear space complexity.
\end{proof}

\section{Counter Example}
\begin{figure}
\begin{minipage}{0.45\textwidth}
\begin{align*}
  g &= \neg (a < b) \vee \neg (b < c) \vee \neg (c < a)\\
    &\wedge (a < b) \vee (b < c) \vee (c < a)\\
    &\wedge \neg (a < b) \vee (b < c) \vee \neg (c < a)\\
    &\wedge \neg (a < b) \vee \neg (b < c) \vee (c < a)\\
    &\wedge \neg (c < b) \vee \neg (c < a)\\
    &\wedge \neg (b < a) \vee \neg (b < c)\\
    &\wedge \neg (a < b) \vee \neg (b < c)\\
    &\wedge \neg (b < a) \vee \neg (c < a)\\
    &\wedge \neg (c < b) \vee \neg (a < b)\\
    &\wedge \neg (c < a) \vee \neg (b < a)\\
\end{align*}
\end{minipage}
\begin{minipage}{0.45\textwidth}
  \begin{align*}
    totality &= (a < b) \vee (b < a)\\
      &\wedge (a < c) \vee (c < a)\\
      &\wedge (b < c) \vee (c < b)\\
      &\wedge \neg(a < b) \vee \neg (b < a)\\
      &\wedge \neg(a < c) \vee \neg (c < a)\\
      &\wedge \neg (b < c) \vee \neg (c < b)\\\\
    transitivity &= (a < b \wedge b < c \Rightarrow a < c)\\
      &\wedge (a < c \wedge c < b \Rightarrow a < b)\\
      &\vdots
  \end{align*}
\end{minipage}
\caption{The formula $g \wedge totality \wedge transitivity$ is a counterexample to Lemma 5.9 of \cite{DBLP:journals/pacmpl/EmmiE18}}
\label{fig:cexformula}
\end{figure}

\begin{figure}
  \includegraphics[width=\linewidth]{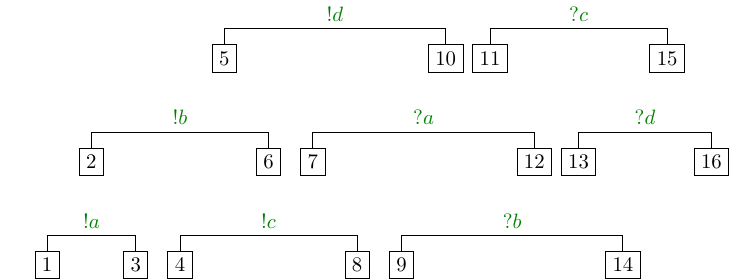}
  \caption{A linearizable stack history. It is incorrectly identified as unlinearizable by the algorithm presented in \cite{DBLP:conf/vmcai/PetersonCD21}.}
  \label{fig:sdhk2}
\end{figure}

\end{document}